\def\eqn#1{\eqno(#1)}

\def\mi{\vspace*{2mm}\noindent}
\def\lbd{\{\hskip-5pt\{}
\def\rbd{\}\hskip-5pt\}}
\def\ladb{\left\langle\phantom{|^|_|}\hskip-12pt\right\langle}
\def\radb{\left\rangle\phantom{|^|_|}\hskip-12pt\right\rangle}
\def\lad{\langle\hskip-3pt\langle}
\def\rad{\rangle\hskip-3pt\rangle}
\def\lb{\{}
\def\rb{\}}
\def\la{\langle}
\def\ra{\rangle}
\def\btau{\mbox{\boldmath$\tau$\unboldmath}}
\def\bxi{\mbox{\boldmath$\xi$\unboldmath}}
\def\bgamma{\mbox{\boldmath$\gamma$\unboldmath}}
\def\bal{\mbox{\boldmath$\alpha$\unboldmath}}
\def\bOm{{\bf\Omega}}
\def\bom{\mbox{\boldmath$\omega$\unboldmath}}
\documentclass[russian,12pt]{article}
\usepackage[T1]{fontenc}
\usepackage[cp866]{inputenc}
\usepackage{babel}
\begin{document}
\oddsidemargin 15mm
{\center

\mi{\bf СЛАБО НЕЛИНЕЙНАЯ УСТОЙЧИВОСТЬ
КОНВЕКТИВНЫХ МАГНИТОГИДРОДИНАМИЧЕСКИХ\\
СИСТЕМ БЕЗ $\alpha-$ЭФФЕКТА\\
К ВОЗМУЩЕНИЯМ С БОЛЬШИМИ МАСШТАБАМИ}

\mi В.А.Желиговский

\mi Международный институт теории прогноза землетрясений\\
и математической геофизики РАН 

\mi Лаборатория общей аэродинамики, Институт механики МГУ\\


}

\medskip Рассмотрена задача о слабо нелинейной устойчивости к возмущениям с
боль\-шими масштабами трехмерных конвективных магнитогидродинамических состо\-яний,
в которых нет $\alpha-$эффекта, или он несущественен (например, из-за наличия
в системе симметрии
относительно центра или вертикальной оси). Предполагается, что в исследуемом
на устойчивость МГД состоянии отсутствуют большие прост\-ранственно-временн\'ые
масштабы, и оно устойчиво к возмущениям с таким же малым пространственным
масштабом, как и в исследуемом состоянии. Выведен\-ные с помощью асимптотических
методов уравнения для средних полей возмуще\-ний обобщают стандартные уравнения
магнитогидродинамики (Навье-Стокса и магнитной индукции). В них
появляются оператор комбинированной вихревой диффузии, вообще говоря,
анизотропный и не обязательно отрицательно опреде\-ленный, и дополнительные
квадратичные члены, аналогичные адвективным.

\mi{\bf 1. Введение.}
Согласно существующим научным представлениям, магнитное поле планет Солнечной
системы с расплавленным ядром, включая Землю, поддер\-живается конвективными
магнитогидродинамическими (МГД) процессами в рас\-плавленном ядре (см., например, Моффат, 1980;
Паркер, 1982; Zeldovich и др., 1983; Паркинсон, 1986; Merrill и др., 1996).
В принципе, эти процессы можно исследовать, численно решая систему описывающих
их уравнений (Навье-Стокса с учетом сил Архимеда, Кориолиса и Лоренца,
магнитной индукции и теплопро\-водности). Этот подход использован, например, в
серии работ Глатцмайера с соавторами: Glatzmaier и Roberts [1995а,б; 1996а,б;
1997а,б], Glatzmaier и др. [1999], Christensen и др. [1999], Olson и др.
[1999], Roberts и Glatzmaier [2001], где удалось воспроизвести дипольную в
главном морфологию магнитного поля Земли и его хаотические инверсии, вычисляя
решения этой системы уравнений в сферическом слое, соответствующем внешнему
ядру.

Однако даже современные компьютеры не позволяют выполнять расчеты
с пространственным и временн\'ым разрешением, достаточным для значений
пара\-метров, характеризующих конвекцию во внешнем ядре Земли. Так, указанная
серия работ Глатцмайера с соавторами сделана для чисел Тейлора и Экмана порядка
$10^3$ и $10^{-6}$, что на порядки величины отличается от их значений для ядра
Земли, $\sim10^9$ и $\sim10^{-9}-10^{-15}$ (оценки по молекулярной или
турбулентной кинематической вязкости), соответственно. Из-за грубости
(для рассмотренных значений параметров) использованного пространственного
разрешения расчеты проводили алгоритмом с численной гипервязкостью, а этот
способ пространствен\-ного сглаживания может существенно искажать результаты
[Zhang и Jones, 1997; Sarson и Jones, 1999; Zhang и Schubert, 2000]: при
использовании гипервязкости неосесимметричные компоненты поля скорости потока
жидкости и магнитного поля недооцениваются по сравнению с осесимметричными, и
дипольное магнитное поле оказывается предпочтительной конфигурацией, тогда как
без ее использова\-ния предпочтительная конфигурация -- квадрупольная [Busse,
2000]. (Кроме того, использование гипервязкости приводит к увеличению жесткости
системы обыкно\-венных дифференциальных уравнений Фурье-Галеркина, к которым
сводится исходная система уравнений после пространст\-венной дискретизации.)
По этим причинам полученное в этой серии работ хорошее качественное
соответствие результатов расчетов с реальным геодинамо следует считать
удивительным [Jones, 2000].

Эта проблема вызвана тем, что геодинамо характеризуется наличием структур,
имеющих иерархию пространственных масштабов, между которыми (как и
в турбулентности вообще) происходит взаимодействие (явления прямого и обратн\-ого
каскада энергии и перемежаемости). Пример таких контрастных структур --
пограничный слой Экмана, возникающий в конвективных потоках вращающейся
жидкости (при условии прилипания на границе), неустойчивость которого может
быть причиной генерации магнитного поля [Ponty и др. 2001а,б, 2003; Rotvig и
Jones, 2002]. Взаимодействие ядра и мантии (core-mantle coupling), которое
считается ответственным за декадную вариацию длины дня (так, размеры
топо\-графических структур на границе раздела ядра и мантии не превышают 5 км;
см., например, Merrill и др., 1996), а также существенная роль, которую играет
турбулентность в процессах генерации, -- другие примеры важности малых
масш\-табов в контексте геофизических приложений.

Даже если конвекция в присутствии магнитного поля рассматривается для
условий Земли, ее необходимо исследовать в целой области в пространстве
пара\-метров, так как реологические соотношения [Christensen, 1989] и значения
пара\-метров [Peltier, 1989], входящие в систему уравнений, определяющую
конвектив\-ную МГД систему, известны только приближенно. (Так, оценки коэффициента
тепловой диффузии в ядре Земли отличаются на несколько порядков, см., напри\-мер,
Merrill и др. [1996].) При этом желательно выявить типичные режимы поведения
системы и локализовать точки бифуркаций, в которых происходит его
перестройка. Это невозможно сделать чисто численно из-за огромного объема
требуемых вычислений; следовательно, необходимо применять аналитические и
гибридные аналитико-вычислительные подходы. Такой подход применен, напри\-мер,
в известных исследованиях геодинамо Брагинского [1964а-д, 1967, 1975] и
Soward [1972, 1974], где с помощью асимптотических разложений была определена
величина коэффициента $\alpha$-эффекта, появляющегося в рассмотренных
ими МГД системах.

Данная статья является непосредственным продолжением серии работ автора, где
был использован такой гибридный подход для анализа линейной [Желиговс\-кий, 2003]
и слабо нелинейной [Желиговский, 2006] устойчивости
МГД простран\-ственно-перио\-дических состояний, а также, совместно с Baptista и
др. [2004, 2005], -- линейной устойчивости стационарных состояний в конвекции
Рэлея-Бенара в слое в присутствии магнитного поля. Характерный пространственный
масштаб возмущений считается су\-щественно больше характерного масштаба
рас\-сматриваемого конвективного МГД состояния. Возмущение раскладывается
в асимптотические степенные ряды по малому параметру $\varepsilon$, отношению
этих масш\-табов. Методы теории осреднения уравнений в частных производных для
много\-масштабных систем (см., например, монографии Bensoussan и др., 1978;
Oleinik и др., 1992; Jikov и др., 1994; Cioranescu и Donato, 1999) позволяют
строго вывести замкнутые уравнения для крупномасштабных структур возмущения, усредненных
по мелким масштабам, в которых влияние мелкомасштабных струк\-тур выражается
через специальные члены с усредненными коэффициентами. Вычисление этих
коэффициентов сводится к численному решению систем линей\-ных дифференциальных
уравнений в частных производных.

В работе Желиговского [2003] предполагалось, что рассматриваемое состояние
стационарно и центрально-симметрично, в работе Желиговского [2006] --
цент\-рально-симметрично, а в работах Baptista и др. [2004, 2005] -- стационарно и
симметрично относительно вертикальной оси. Наличие этих симметрий гаранти\-рует,
что $\alpha$-эффект отсутствует или несущественен. В указанных задачах на
линейную устойчивость главные члены разложений мод МГД возмущений стацио\-нарных
состояний и их инкрементов роста являются соответственно собственными векторами
и собственными значениями оператора комбинированной вихревой (турбулентной)
диффузии. Это оператор в частных производных второго порядка, вообще говоря,
анизотропный и не обязательно знакоопределенный. Если он имеет положительные
собственные значения, говорят, что имеет место явление отрицательной диффузии
[Starr, 1968]. (В кинематическом динамо это явление исследовали Lanotte и др.
[1999], Zheligovsky и др. [2001], Zheligovsky и Podvigina [2003] и Zheligovsky
[2005].) В задаче о слабо нелинейной устойчивости МГД состояний в уравнениях
для средних полей возмущения появляются также дополнительные квадратичные члены,
аналогичные адвективным [Желиговский, 2006]; в этой работе также обобщен метод
Zheligovsky [2005] для экономичного вычисления коэффициентов вихревой диффузии
и адвекции в уравнениях для средних полей. Желиговский [2006] не предполагал
стационарность МГД состоян\-ия, устойчивость которого он исследовал.
В этом случае усреднение необходимо проводить по всей пространст\-венно-временн\'ой
области изменения быстрых пере\-менных, а коэффициенты вихревых тензоров --
константы, в частности, не зави\-сящие от времени; таким образом, исследование
временн\'ой зависимости вихревой вязкости от времени Gama и Chaves
[2000] не имеет под собой математического основания.

Уравнения средних полей для слабо нелинейных возмущений
двумерных ста\-ционарных гидродинамических систем в отсутствие
магнитного поля рассматри\-вали в терминах функции тока Gama и др. [1994]
и Frisch и др. [1996] (в последней работе для учета планетарного вращения
в уравнение Навье-Стокса было включе\-но слагаемое, описывающее $\beta$-эффект,
и была рассмотрена ситуация малой над\-критичности для возникновения эффекта
отрицательной вихревой вязкости).
Уравнения для возмущений, аналогичные по методу построения уравнениям средних
полей, изучали ранее Newell [1983] и Cross и Newell [1984]; в качестве исходных
уравнений, однако, они принимали модельные уравнения, приближенно описывающие
конвективные течения в слое в форме деформированных валов. Newell и др.
[1990а,б, 1993, 1996] и Ponty и др. [1997] исследовали устойчивость системы
конвективных валов и дефекты, возникающие в этой системе, рассмат\-ривая полную
систему уравнений Буссинеска конвекции в слое жидкости с жест\-кими границами
в переменных амплитуда - фаза. Наряду с медленными времен\-н\'ой и горизонтальными
пространственными переменными ими была использована медленная фаза, для которой
выведено усредненное уравнение. В перечисленных выше работах Ньюэлла
с соавторами магнитное поле не рассматривалось.

В этой статье рассмотрена дальнейшая эволюция в слабо нелинейном режиме
возмущений с большими масштабами трехмерного конвективного МГД состояния,
линейную стадию развития которых изучали Baptista и др. [2005], и выведены
уравнения для средних полей возмущений, которые оказываются обобщением
стандартных уравнений магнитогидродинамики.
В отличие от цитированных выше работ Желиговского с соавторами, здесь
рассматривается конвекция в слое, вращающемся вокруг вертикальной оси, что более
естественно для геофизи\-ческих приложений, но приводит к следующей алгебраической
трудности. Стан\-дартный метод вывода уравнений для средних полей возмущений
использует то обстоятельство, что ядро оператора, сопряженного к оператору
линеаризации уравнений в окрестности МГД состояния, устойчивость которого
исследуется, содержит векторные поля - константы. Уравнения для средних полей
возмущения являются условием разрешимости уравнений в быстрых
переменных, отвечающих малым масштабам, которое по теореме Фредгольма
состоит в ортогональности правой части ядру сопряженного оператора и
в рассматриваемом случае эквива\-лентно усреднению уравнений
в быстрых переменных. При наличии силы Корио\-лиса ядро сопряженного
оператора содержит константы, если разрешить линей\-ный рост по горизонтальным
направлениям потенциала вычитаемого градиента. Однако тогда при усреднении
появляется среднее от поверхностного интеграла, соответствующего давлению,
не имеющее вид дифференциального оператора от средних полей возмущений.
Чтобы обойти эту сложность, уравнение Навье-Стокса удобно рассматривать
в форме уравнения для завихренности. Уравнение для усредненного потока
получается тогда как условие разрешимости уравнения при
$\varepsilon^3$, а не $\varepsilon^2$, как обычно.

Выведенные в настоящей статье уравнения для средних полей можно рассма\-тривать
как аналитическое обоснование вычислительного метода крупных вихрей (large eddy
simulations -- см. Sa\-gaut, 2006), если в конвективном МГД состоянии
имеет место разделение масштабов. В этом случае вместо обычно используемых
в этом методе эмпирических формул для оценки влияния мелких масштабов на крупные
и способов замыкания уравнений для средних полей (``крупных вихрей") можно
использовать выведенные здесь точные асимптотические результаты -- вихревые
тензоры, коэффициенты которых выражаются через решения вспомо\-гательных задач.

\pagebreak
\mi{\bf 2. Уравнения конвекции в присутствии магнитного поля.}
Конвективное МГД состояние $\bOm,{\bf V,H},T$, нелинейная устойчивость
которого исследуется, удов\-летворяет системе уравнений

$${\partial\bOm\over\partial t}=\nu\Delta\bOm+\nabla\times({\bf V}\times\bOm
-{\bf H}\times(\nabla\times{\bf H}))+\nabla\times({\bf V\times\tau e}_3
+\beta T{\bf e}_3)+{\bf F},\eqn{1.1}$$
$${\partial{\bf H}\over\partial t}=
\eta\Delta{\bf H}+\nabla\times({\bf V}\times{\bf H})+{\bf J},\eqn{1.2}$$
$${\partial T\over\partial t}=\kappa\Delta T-({\bf V}\cdot\nabla)T+S,$$
$$\nabla\cdot{\bf V}=\nabla\cdot{\bf H}=0,\eqn{1.3}$$
$$\nabla\times{\bf V}=\bOm\eqn{1.4}$$
(принимается приближение Буссинеска). Здесь ${\bf V(x},t)$ -- скорость потока
прово\-дящей жидкости, $\bOm({\bf x},t)$ -- его завихренность, $T({\bf x},t)$
-- температура, ${\bf H(x},t)$ -- магнитное поле, $t$ -- время, $\nu$, $\eta$ и
$\kappa$ -- коэффициенты кинематической, магнитной и тепловой молекулярной
диффузии, соответственно, $\tau/2$ -- скорость вращения слоя жидкости,
$\beta(T-T_0){\bf e}_3$ -- сила Архимеда, ${\bf F(x},t)$ -- объемная сила,
${\bf J(x},t)$ отвечает наличию в системе распределения наложенных внешних
токов, $S({\bf x},t)$ -- источников тепла, ${\bf e}_k$ -- единичный вектор
вдоль оси координат $x_k$.

Горизонтальные поверхности слоя $x_3=\pm L/2$ свободны и элект\-ропроводны:
$$\left.{\partial{\bf V}^1\over\partial x_3}\right|_{x_3=\pm L/2}=
\left.{\partial{\bf V}^2\over\partial x_3}\right|_{x_3=\pm L/2}=0,\quad
\left.{\bf V}^3\right|_{x_3=\pm L/2}=0,\eqn{2.1}$$
$$\Rightarrow\left.\bOm^1\right|_{x_3=\pm L/2}
=\left.\bOm^2\right|_{x_3=\pm L/2}=0,\quad
\left.{\partial\bOm^3\over\partial x_3}\right|_{x_3=\pm L/2}=0;\eqn{2.2}$$
$$\left.{\partial{\bf H}^1\over\partial x_3}\right|_{x_3=\pm L/2}=
\left.{\partial{\bf H}^2\over\partial x_3}\right|_{x_3=\pm L/2}=0,\quad
\left.{\bf H}^3\right|_{x_3=\pm L/2}=0;\eqn{2.3}$$
$$\left.\phantom{|_|}T\right|_{x_3=-L/2}=T_1,\quad
\left.\phantom{|_|}T\right|_{x_3=L/2}=T_2$$
(верхний индекс нумерует компоненты вектора). Конвекция возможна только при
$T_1>T_2$. Удобно ввести новую переменную $\Theta=T-T_1+\delta(x_3+L/2)$,
где $\delta=(T_1-T_2)/L$, удовлетворяющую уравнению
$${\partial\Theta\over\partial t}=\kappa\Delta\Theta
-({\bf V}\cdot\nabla)\Theta-\delta{\bf V}^3+S\eqn{3}$$
и краевым условиям
$$\Theta\left|_{x_3=\pm L/2}=0.\right.\eqn{4}$$

Для системы в слабо нелинейном режиме считаем, что амплитуда возмущения
порядка $\varepsilon$. Возмущенное состояние \hbox{$\bf V+\varepsilon v$,}
$\bOm+\varepsilon\bom,{\bf H+\varepsilon h},\Theta+\varepsilon\theta$
также удовлетво\-ряет системе (1)-(4), откуда профили возмущений
$\bom,{\bf v,h},\theta$ (в дальнейшем эти поля
будем называть просто возмущениями) удовлетворяют уравнениям
$${\partial\bom\over\partial t}=\nu\Delta\bom+\nabla\times({\bf V}\times\bom
+{\bf v}\times\bOm-{\bf H}\times(\nabla\times{\bf h})
-{\bf h}\times(\nabla\times{\bf H}))$$
$$+\tau{\partial{\bf v}\over\partial x_3}+\beta\nabla\theta\times{\bf e}_3
+\varepsilon\nabla\times({\bf v}\times\bom-{\bf h}\times(\nabla\times{\bf h})),\eqn{5.1}$$
$${\partial{\bf h}\over\partial t}=\eta\Delta{\bf h}
+\nabla\times({\bf v\times H+V\times h+\varepsilon v\times h}),\eqn{5.2}$$
$${\partial\theta\over\partial t}=\kappa\Delta\theta
-({\bf V}\cdot\nabla)\theta-({\bf v}\cdot\nabla)\Theta
-\varepsilon({\bf v}\cdot\nabla)\theta-\delta{\bf v}^3,\eqn{5.3}$$
$$\nabla\cdot{\bf v}=\nabla\cdot{\bf h}=0,\eqn{5.4}$$
$$\nabla\times{\bf v}=\bom.\eqn{5.5}$$

Усреднение по объему слоя горизонтальных компонент уравнения Навье-Стокса
(5.1) для возмущения, представленного в форме для скорости потока:
$${\partial{\bf v}\over\partial t}=\nu\Delta{\bf v}+{\bf V}\times\bom
+{\bf v}\times\bOm-{\bf H}\times(\nabla\times{\bf h})
-{\bf h}\times(\nabla\times{\bf H})$$
$$+\tau{\bf v}\times{\bf e}_3+\beta\theta{\bf e}_3
+\varepsilon({\bf v}\times\bom-{\bf h}\times(\nabla\times{\bf h}))-\nabla p,$$
дает
$${\partial\la{\bf v}\ra_h\over\partial t}=
\la{\bf v}\ra_h\times\tau{\bf e}_3-\la\nabla p\ra_h,$$
где обозначено
$$\la{\bf f}\ra_h\equiv\la{\bf f}^1\ra{\bf e}_1+\la{\bf f}^2\ra{\bf e}_2,$$
$$\la f({\bf x},t)\ra\equiv\lim_{\ell\to\infty}{1\over L\ell^2}\int_{-L/2}^{L/2}
\int_{-\ell/2}^{\ell/2}\int_{-\ell/2}^{\ell/2}f({\bf x},t)\,dx_1\,dx_2\,dx_3.$$
Таким образом, значениями среднего горизонтальной компоненты возмущения скорости
потока управляет среднее градиента давления $p$, которое должно быть задано.
Мы рассматриваем случай, когда прокачивания жидкости сквозь слой за счет
перепада давления на бесконечности не происходит; это имеет место,
если рост давления на бесконечности ограничен так, что $\la\nabla p\ra_h=0$.
Тогда, если в начальный момент горизонтальная компонента возмущения
скорости равна нулю, то и в любой момент
$$\la{\bf v}\ra_h=0.\eqn{5.6}$$

\mi{\bf 3. Предположения об исследуемом конвективном МГД состоянии}
$\bOm,\bf V$, ${\bf H},\Theta$.
Мы считаем, что выполнены следующие условия:

1) Поля $\bOm,{\bf V,H},\Theta$,
определяющие исходное конвективное МГД состояние, устойчивость которого
исследуется, гладки и глобально ограничены, и корректны все
пространственно-временн\'ые усреднения по быстрым переменным, проводимые в
процессе вывода уравнений средних полей. Эти условия, как легко видеть, выполнены, если,
например, исходное состояние периодично по простран\-ству и стационарно или
периодично по времени.

Пространственное среднее $f$ по быстрым переменным определяется равенством
$$\la f\ra\equiv\lim_{\ell\to\infty}{1\over L\ell^2}\int_{-L/2}^{L/2}
\int_{-\ell/2}^{\ell/2}\int_{-\ell/2}^{\ell/2}f({\bf x},...)\,dx_1\,dx_2\,dx_3,$$
пространственно-временн\'ое среднее по быстрым переменным и осциллирующая
часть поля $f$, -- соответственно,
$$\lad f\rad\equiv\lim_{\hat{t}\to\infty}
{1\over\hat{t}}\int_0^{\hat{t}}\la f({\bf x},t,{\bf X},T)\ra\,dt,\qquad
\lbd f\rbd\equiv f-\lad f\rad.$$
Обозначим также
$$\la{\bf f}\ra_v\equiv\la{\bf f}^3\ra{\bf e}_3,\qquad
\lb{\bf f}\rb_v\equiv{\bf f}-\la{\bf f}\ra_v;$$
$$\lad{\bf f}\rad_v\equiv\lad{\bf f}^3\rad{\bf e}_3,\qquad
\lbd{\bf f}\rbd_v\equiv{\bf f}-\lad{\bf f}\rad_v;$$
$$\la{\bf f}\ra_h\equiv\la{\bf f}^1\ra{\bf e}_1+\la{\bf f}^2\ra{\bf e}_2,\qquad
\lb{\bf f}\rb_h\equiv{\bf f}-\la{\bf f}\ra_h;$$
$$\lad{\bf f}\rad_h\equiv\lad{\bf f}^1\rad{\bf e}_1+\lad{\bf f}^2\rad{\bf e}_2,\qquad
\lbd{\bf f}\rbd_h\equiv{\bf f}-\lad{\bf f}\rad_h.$$

2) В системе отсутствует или несущественен $\alpha-$эффект. Это условие
будет обсуждено детально при рассмотрении решения уравнений порядка
$\varepsilon^1$ и $\varepsilon^2$. Как будет показано, оно заведомо выполнено,
если исходное конвективное МГД состо\-яние центрально-симметрично:
$${\bf V}(-{\bf x},t)=-{\bf V}({\bf x},t)\quad\Rightarrow\quad
\bOm(-{\bf x},t)=\bOm({\bf x},t);$$
$${\bf H}(-{\bf x},t)=-{\bf H}({\bf x},t);\quad\Theta(-{\bf x},t)=-\Theta({\bf x},t)$$
(т.е. $\bf V$ и $\bf H$ центрально-симметричны, а $\bOm$ и $\theta$
центрально-анти\-симметричны),
или если оно симметрично относительно вертикальной оси:
$${\bf V}^1(-x_1,-x_2,x_3)=-{\bf V}^1(x_1,x_2,x_3),$$
$${\bf V}^2(-x_1,-x_2,x_3)=-{\bf V}^2(x_1,x_2,x_3),$$
$${\bf V}^3(-x_1,-x_2,x_3)={\bf V}^3(x_1,x_2,x_3),$$
такие же соотношения выполнены для $\bOm$ и $\bf H$,
$$\Theta(-x_1,-x_2,x_3)=\Theta(x_1,x_2,x_3).$$
Эти два типа симметрий совместимы с уравнениями (1), (3) (когда выполнены
соответствующие условия для источниковых членов $\bf F,J$ и $S$) и краевыми
услови\-ями (2), (4). Отметим, что наличие симметрий какого-либо из этих типов не
необходимо для отсутствия $\alpha-$эффекта (например, кинематический
$\alpha-$эффект не появляется в ABC-потоках [Wirth и др., 1995]).

Мы называем векторное поле $\bf f$ антисимметричным, если
$${\bf f}(-{\bf x},t)={\bf f}({\bf x},t)$$
для случая центральной симметрии, или если
$${\bf f}^1(-x_1,-x_2,x_3)={\bf f}^1(x_1,x_2,x_3),$$
$${\bf f}^2(-x_1,-x_2,x_3)={\bf f}^2(x_1,x_2,x_3),$$
$${\bf f}^3(-x_1,-x_2,x_3)=-{\bf f}^3(x_1,x_2,x_3)$$
для случая симметрии относительно вертикальной оси. Скалярное поле $f$
анти\-симметрично, если
$$f(-{\bf x},t)=f({\bf x},t),$$
или
$$f(-x_1,-x_2,x_3)=-f(x_1,x_2,x_3),$$
соответственно.

3) Пространственно-временн\'ые средние горизонтальных компонент полей $\bf V$
и $\bf H$ равны нулю. Это условие, очевидно, выполнено
автоматически, если исходное состояние центрально-симмет\-рично или симметрично
относительно вертикальной оси.

4) Исходные состояния устойчивы к возмущениям с малым
про\-странственным масштабом. Операторы линеаризации исходных уравнений (5)
в окрестности исследуемого конвективного МГД состояния имеют вид
$${\cal L}^\omega(\bom,{\bf v,h},\theta)\equiv\left.-{\partial\bom\over\partial t}
+\nu\Delta_{\bf x}\bom+\nabla_{\bf x}\times\right(\bf V\times\bom+v\times\bOm$$
$${\bf-H\times(\nabla_x\times h)-h\times(\nabla_x\times H})\left)
+\tau{\partial{\bf v}\over\partial x_3}\right.+\beta\nabla_{\bf x}\theta\times{\bf e}_3,$$
$${\cal L}^h({\bf v,h})\equiv-{\partial{\bf h}\over\partial t}+\eta\Delta_{\bf x}{\bf h}
+\nabla_{\bf x}\times({\bf v\times H+V\times h}),$$
$${\cal L}^\theta({\bf v},\theta)\equiv-{\partial\theta\over\partial t}+\kappa\Delta_{\bf x}\theta
-({\bf V}\cdot\nabla_{\bf x})\theta-({\bf v}\cdot\nabla_{\bf x})\Theta-\delta{\bf v}^3,$$
где
$$\nabla_{\bf x}\cdot\bom=\nabla_{\bf x}\cdot{\bf h}=0,$$
и ${\bf v(x},t)$ удовлетворяет системе уравнений
$$\nabla_{\bf x}\times{\bf v}=\bom,\qquad\nabla_{\bf x}\cdot{\bf v}=0,\qquad
\la{\bf v}\ra_h=0.\eqn{6}$$
Оператор линеаризации для уравнения Навье-Стокса в форме для скорости потока,
эквивалентный ${\cal L}^\omega$, имеет вид
\pagebreak
$${\cal L}^v({\bf v,h},\theta,p)\equiv-{\partial{\bf v}\over\partial t}
+\nu\Delta_{\bf x}{\bf v}+{\bf V}\times(\nabla_{\bf x}\times{\bf v})
+{\bf v}\times\bOm$$
$$-{\bf H}\times(\nabla_{\bf x}\times{\bf h})
-{\bf h}\times(\nabla_{\bf x}\times{\bf H})
+\tau{\bf v}\times{\bf e}_3+\beta\theta{\bf e}_3-\nabla_{\bf x}p.$$

Пусть $\bom,\bf v,h$ и $\theta$ -- произвольные глобально ограниченные вместе
с их производ\-ными поля, удовлетворяющие граничным условиям (2) и (4).
Поскольку $\bf V$ и $\bf H$ принадлежат этому классу,
$$\la{\cal L}^\omega(\bom,{\bf v,h},\theta)\ra_v=-{\partial\la\bom\ra_v\over\partial t};\eqn{7.1}$$
$$\la{\cal L}^h({\bf v,h})\ra_h=-{\partial\la{\bf h}\ra_h\over\partial t};\eqn{7.2}$$
$$\Rightarrow\quad\lad{\cal L}^\omega(\bom,{\bf v,h},\theta)\rad_v=0,\qquad
\lad{\cal L}^h({\bf v,h})\rad_h=0.\eqn{7.3}$$
Из (7.3) следует, что условия
$$\lad{\bf f}^\omega({\bf x},t)\rad_v=0,\eqn{8.1}$$
$$\lad{\bf f}^h({\bf x},t)\rad_h=0\eqn{8.2}$$
необходимы для существования решений системы уравнений
$${\cal L}^\omega(\bom,{\bf v,h},\theta)={\bf f}^\omega,\qquad{\cal L}^h({\bf v,h})={\bf f}^h,
\qquad{\cal L}^\theta({\bf v},\theta)=f^\theta,\eqn{9.1}$$
$$\nabla_{\bf x}\cdot\bom=\nabla_{\bf x}\cdot{\bf h}=0\eqn{9.2}$$
(с учетом (6)) из рассматриваемого класса. Вследствие (7.1) и (7.2) пространст\-венные
средние $\la\bom\ra_v$ и $\la{\bf h}\ra_h$ решений системы
$${\cal L}^\omega(\bom,{\bf v,h},\theta)=0,\quad{\cal L}^h({\bf v,h})=0,\quad{\cal L}^\theta({\bf v},\theta)=0,\eqn{10.1}$$
$$\nabla_{\bf x}\cdot\bom=\nabla_{\bf x}\cdot{\bf h}=0\eqn{10.2}$$
из этого класса функций сохраняются во времени, поэтому линейная устойчивость
исходного конвективного МГД состояния не асимптотическая. Будем считать,
что всякое решение (10),(6) из указанного класса функций с нулевым
средними $\la\bom\ra_v$ и $\la{\bf h}\ra_h$ экспоненциально затухает во времени.

5) Для произвольных глобально ограниченных вместе с производными глад\-ких
соленоидальных полей ${\bf f}^\omega({\bf x},t)$, ${\bf f}^h({\bf x},t)$ и
$f^\theta({\bf x},t)$ с нулевыми средними (8) система (9),(6) имеет при заданных
начальных условиях единственное решение $\bom,{\bf v,h},\theta$ в указанном пространстве.
Это условие гарантирует, что имеют решение т.н. вспомогательные задачи,
сформулированные ниже. Оно выполнено, например, для пространственно-периодических
стационарных или периодических по време\-ни устойчивых к короткомасштабным
(не зависящим от медленных переменных) возмущениям состояний конвективных МГД
систем общего положения (малое возмущение полей $\bf F,J$ и $S$ в системах, где
оно не выполнено, приводит либо к неустойчивости, либо к его выполнению).
Решение (9),(6) может быть построено, как решение параболического уравнения
(однако поскольку область, которую занимает объем жидкости,
не компактна, не гарантировано, что оно будет гло\-бально ограничено
вместе с производными). Его единственность при данных начальных условиях
следует из сделанного выше предположения о характере линейной устойчивости
исходного конвективного МГД состояния.

Мы не фиксируем какие-либо краевые условия для возмущенного состояния
или усредненного возмущения, для которого мы выводим замкнутые уравнения.
Выведенные уравнения средних полей описывают физически реализуемое состоя\-ние,
если их решение глобально ограничено по пространству (при этом оно приме\-нимо
на тех отрезках медленного времени, на которых оно остается глобально
ограниченным). Таким образом, единственное краевое условие состоит в требова\-нии
глобальной ограниченности возмущения по пространству в каждый момент времени.
В расчетах его естественно заменить на более жесткое условие простран\-ственной
периодичности или квазипериодичности усредненного возмущения по медленным
переменным.

\mi{\bf 4. Формальные асимптотические разложения.}
Следуя стандартным мето\-дам осреднения уравнений в частных производных, вводим
быстрые пространст\-венные $\bf x$ и временн\'ую $t$ переменные и соответствующие
медленные переменные ${\bf X}=\varepsilon(x_1,x_2),T=\varepsilon^2t$.
Ищем решение задачи (5) в виде формальных степенных рядов
$$\bom=\sum_{n=0}^\infty\bom_n({\bf x},t,{\bf X},T)\varepsilon^n,\eqn{11.1}$$
$${\bf v}=\sum_{n=0}^\infty{\bf v}_n({\bf x},t,{\bf X},T)\varepsilon^n,\eqn{11.2}$$
$${\bf h}=\sum_{n=0}^\infty{\bf h}_n({\bf x},t,{\bf X},T)\varepsilon^n,\eqn{11.3}$$
$$\theta=\sum_{n=0}^\infty\theta_n({\bf x},t,{\bf X},T)\varepsilon^n\eqn{11.4}$$
(считаем, что в начальный момент времени все члены рядов (11) заданы).
Рассмат\-ривая осредненную по быстрым переменным и осциллирующую части полученной
иерархии уравнений последовательно при каждой степени $\varepsilon^n$, мы
выведем (как условие разрешимости уравнений в быстрых переменных при $n=2$ и 3)
замкнутые нелинейные уравнения в медленных переменных для усредненного главного
члена разложения возмущения (уравнения средних полей).

Приравняем нулю коэффициенты рядов по степеням $\varepsilon$, полученных
подстанов\-кой рядов (11.2) и (11.3) в (1.3).
Выделяя среднюю и осциллирующую часть полученных уравнений, находим
$$\nabla_{\bf X}\cdot\la{\bf v}_n\ra_h=\nabla_{\bf X}\cdot\la{\bf h}_n\ra_h=0,\eqn{12.1}$$
$$\nabla_{\bf x}\cdot\lb{\bf v}_n\rb_h+\nabla_{\bf X}\cdot\lb{\bf v}_{n-1}\rb_h=0,\eqn{12.2}$$
$$\nabla_{\bf x}\cdot\lb{\bf h}_n\rb_h+\nabla_{\bf X}\cdot\lb{\bf h}_{n-1}\rb_h=0.\eqn{12.3}$$
Очевидно,
$$\nabla_{\bf X}\cdot\la\bom_n\ra_v=0,\eqn{12.4}$$
и из уравнения $\nabla\cdot\bom=0$
$$\nabla_{\bf x}\cdot\lb\bom_n\rb_v+\nabla_{\bf X}\cdot\lb\bom_{n-1}\rb_v=0\eqn{12.5}$$
при всех $n\ge0$. (В дифференциальных операторах с индексами
$\bf x$ и $\bf X$ дифферен\-цирование производится по быстрым и медленным
пространственным перемен\-ным, соответственно; все члены разложений с индексом
$n<0$ по определению равны 0.)

Очевидно, $\la\bom_n\ra_v$, $\la{\bf v}_n\ra_h$ и $\la{\bf h}_n\ra_h$
удовлетворяют краевым условиям (2). В силу (12.1) $\la{\bf v}_n\ra_h$
выражается через функцию тока $\psi_n(X_1,X_2,t,T)$:
$$\la{\bf v}_n\ra_h=\left(-{\partial\psi_n\over\partial X_2},
{\partial\psi_n\over\partial X_1},0\right).$$
Подставив ряды (11.1) и (11.2) в (1.4), находим
$$\nabla_{\bf x}\times\lb{\bf v}_n\rb_h=\bom_n-\nabla_{\bf X}\times{\bf v}_{n-1}.\eqn{13}$$
Это равенство -- уравнение относительно $\lb{\bf v}_n\rb_h$. Рассмотрим
оператор ротор, действующий из пространства соленоидальных полей,
удовлетворяющих краевым условиям (2.1), в пространство соленоидальных
полей. Сопряженный к нему оператор, также ротор, определен на пространстве соленоидальных
полей, удов\-летворяющих краевым условиям (2.1). Ядро сопряженного
оператора состоит из постоянных векторных полей вида $(0,0,C)$. Легко показать
по индукции, используя (12.5), что правая часть (13) соленоидальна в быстрых переменных.
Соответственно, (13) разрешимо, только если
$$\la\bom_n\ra^3=(\nabla_{\bf X}\times\la{\bf v}_{n-1}\ra)\cdot{\bf e}_3
\quad\Leftrightarrow\quad\la\bom_n\ra_v=\nabla_{\bf X}\times\la{\bf v}_{n-1}\ra_h.\eqn{14}$$
(При $n=0$ отсюда $\la\bom_0\ra_v=0$.) В терминах функции тока это уравнение
имеет вид $\Delta_{\bf X}\psi_{n-1}=\la\bom_n\ra^3$; отсюда можно определить
$\la{\bf v}_{n-1}\ra_h$, используя условие глобаль\-ной ограниченности
$\psi_{n-1}$ (для выполнения условия (5.6) при усреднении потока по медленным
переменным), если среднее от $\la\bom_n\ra^3$ по плоскости
медленных перемен\-ных равно 0.

Решение (12.2) и (13) c учетом (14) ищем в виде
$$\lb{\bf v}_n\rb_h={\bf A}+\nabla_{\bf x}B,$$
где $\bf A$ -- решение уравнения Пуассона
$$\Delta_{\bf x}{\bf A}=-\nabla_{\bf x}\times
(\lb\bom_n\rb_v-\nabla_{\bf X}\times\lb{\bf v}_{n-1}\rb_h),$$
$$\left.{\partial{\bf A}^1\over\partial x_3}\right|_{x_3=\pm L/2}=
\left.{\partial{\bf A}^2\over\partial x_3}\right|_{x_3=\pm L/2}=0,\quad
\left.{\bf A}^3\right|_{x_3=\pm L/2}=0,$$
а $B$ -- решение задачи Неймана
$$\Delta_{\bf x}B=-\nabla_{\bf X}\cdot\lb{\bf v}_{n-1}\rb_h
-\nabla_{\bf x}\cdot{\bf A},\qquad
\left.{\partial B\over\partial x_3}\right|_{x_3=\pm L/2}=0.$$

Подставив (11) в уравнения (5.1)-(5.3), преобразуем их к виду равенств
рядов по степеням $\varepsilon$. Приравнивая коэффициенты этих рядов, получаем
рекуррентную систему уравнений, которую последовательно решаем совместно с
(12),(13), выде\-ляя среднюю и осциллирующую часть каждого уравнения.

\mi{\bf 5. Уравнения порядка $\varepsilon^0$.}
Из главных членов рядов (5.1)-(5.3) получаем уравнения
$${\cal L}^\omega(\bom_0,{\bf v}_0,{\bf h}_0,\theta_0)=0\eqn{15.1}$$
$$\Leftrightarrow\quad{\cal L}^\omega(\lbd\bom_0\rbd_v,\lbd{\bf v}_0\rbd_h,\lbd{\bf h}_0\rbd_h,\theta_0)
+\nabla_{\bf x}\times(\lad{\bf v}_0\rad_h\times\bOm
-\lad{\bf h}_0\rad_h\times(\nabla_{\bf x}\times{\bf H}))=0;$$
$${\cal L}^h({\bf v}_0,{\bf h}_0)=0\eqn{15.2}$$
$$\Leftrightarrow\quad{\cal L}^h(\lbd{\bf v}_0\rbd_h,\lbd{\bf h}_0\rbd_h)+(\lad{\bf h}_0\rad_h\cdot\nabla_{\bf x}){\bf V}
-(\lad{\bf v}_0\rad_h\cdot\nabla_{\bf x}){\bf H}=0;$$
$${\cal L}^\theta({\bf v}_0,\theta_0)=0\eqn{15.3}$$
$$\Leftrightarrow\quad{\cal L}^\theta(\lbd{\bf v}_0\rbd_h,\theta_0)
-(\lad{\bf v}_0\rad_h\cdot\nabla_{\bf x})\Theta=0.$$
$\lad{\bf v}_0\rad_h$ и $\lad{\bf h}_0\rad_h$
не зависят от быстрых переменных, а в операторах $\cal L$ дифференци\-рование
проводится только по быстрым переменным, поэтому в силу линейности
уравнения (15), (12) и (13) при $n=0$ имеют решения следующей структуры:
$$\lbd\bom_0\rbd_v=\bxi_0^\omega+\sum_{k=1}^2\left({\bf S}^{v,\omega}_k\lad{\bf v}^k_0\rad
+{\bf S}^{h,\omega}_k\lad{\bf h}^k_0\rad\right),\eqn{16.1}$$
$$\lbd{\bf v}_0\rbd_h=\bxi_0^v+\sum_{k=1}^2\left({\bf S}^{v,v}_k\lad{\bf v}^k_0\rad
+{\bf S}^{h,v}_k\lad{\bf h}^k_0\rad\right),\eqn{16.2}$$
$$\lbd{\bf h}_0\rbd_h=\bxi_0^h+\sum_{k=1}^2\left({\bf S}^{v,h}_k\lad{\bf v}^k_0\rad
+{\bf S}^{h,h}_k\lad{\bf h}^k_0\rad\right),\eqn{16.3}$$
$$\theta_0=\xi_0^\theta+\sum_{k=1}^2\left(S^{v,\theta}_k\lad{\bf v}^k_0\rad
+S^{h,\theta}_k\lad{\bf h}^k_0\rad\right).\eqn{16.4}$$
При $n=0$ (13) принимает вид $\nabla_{\bf x}\times{\bf v}_0=\bom_0$, откуда
${\bf S}^{\cdot,v}_k$ определяются через ${\bf S}^{\cdot,\omega}_k$ из уравнений
$$\nabla_{\bf x}\times{\bf S}^{v,v}_k={\bf S}^{v,\omega}_k,\quad
\nabla_{\bf x}\times{\bf S}^{h,v}_k={\bf S}^{h,\omega}_k,\eqn{17.1}$$
$$\nabla_{\bf x}\cdot{\bf S}^{v,v}_k=\nabla_{\bf x}\cdot{\bf S}^{h,v}_k=0.\eqn{17.2}$$

Функции ${\bf S}^{\cdot,\cdot}_k({\bf x},t)$ удовлетворяют
краевым условиям вида (2) и (4), и являются решениями
{\it вспомогательных задач первого типа}
\pagebreak
$${\cal L}^\omega({\bf S}^{v,\omega}_k,{\bf S}^{v,v}_k,{\bf S}^{v,h}_k,S^{v,\theta}_k)=
{\partial\bOm\over\partial x_k},\eqn{18.1}$$
$$\nabla_{\bf x}\cdot{\bf S}^{v,\omega}_k=0,\eqn{18.2}$$
$${\cal L}^h({\bf S}^{v,v}_k,{\bf S}^{v,h}_k)=
{\partial{\bf H}\over\partial x_k},\eqn{18.3}$$
$$\nabla_{\bf x}\cdot{\bf S}^{v,h}_k=0,\eqn{18.4}$$
$${\cal L}^\theta({\bf S}^{v,v}_k,S^{v,\theta}_k)=
{\partial\Theta\over\partial x_k};\eqn{18.5}$$
$${\cal L}^\omega({\bf S}^{h,\omega}_k,{\bf S}^{h,v}_k,{\bf S}^{h,h}_k,S^{h,\theta}_k)=
-{\partial\over\partial x_k}(\nabla_{\bf x}\times{\bf H}),\eqn{19.1}$$
$$\nabla_{\bf x}\cdot{\bf S}^{h,\omega}_k=0,\eqn{19.2}$$
$${\cal L}^h({\bf S}^{h,v}_k,{\bf S}^{h,h}_k)=
-{\partial{\bf V}\over\partial x_k},\eqn{19.3}$$
$$\nabla_{\bf x}\cdot{\bf S}^{h,h}_k=0,\eqn{19.4}$$
$${\cal L}^\theta({\bf S}^{h,v}_k,S^{h,\theta}_k)=0\eqn{19.5}$$
совместно с (17). Векторные потенциалы (18.1) и (19.1) представляют собой
замкнутые уравнения для поля скорости потока:
$${\cal L}^v({\bf S}^{v,v}_k,{\bf S}^{v,h}_k,S^{v,\theta}_k,S^{v,p}_k)=
{\partial{\bf V}\over\partial x_k};$$
$${\cal L}^v({\bf S}^{h,v}_k,{\bf S}^{h,h}_k,S^{h,\theta}_k,S^{h,p}_k)=
-{\partial{\bf H}\over\partial x_k},$$
где $\la\nabla_{\bf x}S^{\cdot,p}_k\ra_h=0$.

Усредняя вертикальные компоненты (18.1) и (19.1) по слою, получаем
$${\partial\la{\bf S}^{\cdot,\omega}_k\ra_v\over\partial t}=0,$$
откуда условия разрешимости для уравнений (17), $\la{\bf S}^{\cdot,\omega}_k\ra_v=0$,
(ср. (14)) выполне\-ны в любой момент времени, если они выполнены при $t=0$.
Аналогично, сохраня\-ется во времени $\la{\bf S}^{\cdot,h}_k\ra_h$. Взяв
дивергенцию уравнений (18.1), (18.3), (19.1) и (19.3), находим, что
${\bf S}^{\cdot,\cdot}_k$ соленоидальны при $t>0$ тогда и только тогда, когда
они соленоидальны при $t=0$. Условия (18.2), (18.4), (19.2) и (19.4)
необходимы для выполнения (12.3) и (12.5) при $n=0$. Таким образом,
начальные условия для задач (18) и (19) --
глобально ограниченные вместе с производными гладкие поля,
удовлетворяющие (18.2), (18.4), (19.2), (19.4),
$\la{\bf S}^{\cdot,\omega}_k\ra_v=0$ и $\la{\bf S}^{\cdot,h}_k\ra_h=0$.
Тогда (18) и (19) имеют единственные
решения согласно предположению о разрешимости задач (9) (условие разрешимости
(8) для них выполнено в силу глобальной ограни\-ченности полей $\bOm$ и~$\bf H$
и их производных).

Если исходное МГД состояние центрально-симметрично, то про\-странства
цент\-рально-симметричных и центрально-антисиммет\-ричных состояний инвариантны
для оператора ${\cal L}=({\cal L}^\omega,{\cal L}^h,{\cal L}^\theta)$,
поэтому ${\bf S}^{\cdot,\omega}_k$ и $S^{\cdot,\theta}_k$
центрально-симметричны, а ${\bf S}^{\cdot,v}_k$ и ${\bf S}^{\cdot,h}_k$
централь\-но-антисимметричны (если начальные условия для этих полей обладают
соответствующими симметриями). Если исходное МГД состояние сим\-метрично
относительно вертикальной оси, то тип симметрии состояния также меняется на
противоположный: ${\bf S}^{\cdot,\omega}_k$, ${\bf S}^{\cdot,v}_k$,
${\bf S}^{\cdot,h}_k$ и $S^{\cdot,\theta}_k$ антисимметричны
относитель\-но вертикальной оси (если начальные условия антисимметричны).
Если исходное МГД состояние имеет симметрию относительно центра или
вертикальной оси, то мы требуем, чтобы начальные условия для
${\bf S}^{\cdot,\omega}_k$ имели указанные типы симметрий.

$\bxi_0({\bf x},t,{\bf X},T)$ удовлетворяет уравнениям
$${\cal L}^\omega(\bxi_0^\omega,\bxi_0^v,\bxi_0^h,\xi_0^\theta)={\cal L}^h(\bxi_0^v,\bxi_0^h)=0,\quad
{\cal L}^\theta(\bxi_0^v,\xi_0^\theta)=0,\eqn{20.1}$$
$$\nabla_{\bf x}\cdot\bxi_0^\omega=\nabla_{\bf x}\cdot\bxi_0^v=\nabla_{\bf x}\cdot\bxi_0^h=0,\eqn{20.2}$$
$$\nabla_{\bf x}\times\bxi_0^v=\bxi_0^\omega,\eqn{20.3}$$
$$\la\bxi_0^\omega\ra_v=0,\qquad\la\bxi_0^h\ra_h=0.\eqn{20.4}$$
Как и в задачах (18) и (19), соленоидальность $\bxi_0^\cdot$ и выполнение условий (20.4) достаточно
требовать при $t=0$. Начальные условия для задачи (20) выбираем из представлений
$$\bom_0=\bxi_0^\omega+\sum_{k=1}^2\left({\bf S}^{v,\omega}_k\lad{\bf v}^k_0\rad
+{\bf S}^{h,\omega}_k\lad{\bf h}^k_0\rad\right),\eqn{21.1}$$
$${\bf h}_0=\bxi_0^h+\sum_{k=1}^2\left({\bf S}^{v,h}_k\lad{\bf v}^k_0\rad
+{\bf S}^{h,h}_k\lad{\bf h}^k_0\rad\right)+\lad{\bf h}_0\rad_h,\eqn{21.2}$$
$$\theta_0=\xi_0^\theta+\sum_{k=1}^2\left(S^{v,\theta}_k\lad{\bf v}^k_0\rad
+S^{h,\theta}_k\lad{\bf h}^k_0\rad\right)\eqn{21.3}$$
(ср. (16)) при $t=0$. (Усреднение горизонтальных компонент (21.2) по быстрым
пространственным переменным дает
$$\lad{\bf h}_0\rad_h|_{T=0}=\la{\bf h}_0\ra_h|_{t=0};$$
вычтя это равенство из (21.2), находим начальные условия для $\bxi_0^h$.)
Изменение начальных условий для ${\bf S}^{\cdot,\cdot}_k$ в рассматриваемом
классе начальных условий компен\-сируется соответствующим изменением начальных
условий для $\bxi^\cdot_0$, однако, как показано ниже, эта неоднозначность не влияет на вид
уравнений для средних полей, поскольку $\bxi_0$ и эти изменения ${\bf S}^{\cdot,\cdot}_k$
экспоненциально затухают во времени (т.к. они являются решениями
задачи (10) с нулевыми средними $\la\bxi_0^\omega\ra_v=0$,
$\la\bxi_0^h\ra_h=0$). Если исходное конвективное МГД состояние стационарно
или перио\-дично по быстрому времени, для удобства вычисления
пространственно-времен\-н\'ых средних естественно требовать, чтобы функции
${\bf S}^{\cdot,\cdot}_k$ были, соответственно, стационарными или периодичными
по времени решениями задач (18) и (19). Существование стационарных решений
при рассматриваемых условиях можно доказать аналогично работе [Желиговский, 2003],
периодических по времени -- [Zheligovsky и Podvigina, 2003]. Тогда $\bxi_0$
имеет смысл затухающих переходных процессов, приводящих к установившемуся (в
быстром времени) режиму.

\mi{\bf 6. Уравнения порядка $\varepsilon^1$.}
Уравнения, полученные из членов рядов (5.1)-(5.3) порядка $\varepsilon$, имеют вид
$${\cal L}^\omega(\bom_1,{\bf v}_1,{\bf h}_1,\theta_1)
+2\nu(\nabla_{\bf x}\cdot\nabla_{\bf X})\lbd\bom_0\rbd_v
-\nabla_{\bf x}\times({\bf H}\times(\nabla_{\bf X}\times{\bf h}_0))$$
$$+\nabla_{\bf X}\times\left({\bf V}\times\bom_0+{\bf v}_0\times\bOm
-{\bf H\times(\nabla_x\times h}_0)-{\bf h}_0\times(\nabla_{\bf x}\times{\bf H})\right)$$
$$+\nabla_{\bf x}\times({\bf v}_0\times\bom_0-{\bf h}_0\times(\nabla_{\bf x}\times{\bf h}_0))
+\beta\nabla_{\bf X}\theta_0\times{\bf e}_3=0;\eqn{22.1}$$
$${\cal L}^h({\bf v}_1,{\bf h}_1)
+2\eta(\nabla_{\bf x}\cdot\nabla_{\bf X})\lbd{\bf h}_0\rbd_h$$
$$+\nabla_{\bf X}\times({\bf v}_0\times{\bf H}+{\bf V}\times{\bf h}_0)
+\nabla_{\bf x}\times({\bf v}_0\times{\bf h}_0)=0;\eqn{22.2}$$
$${\cal L}^\theta({\bf v}_1,\theta_1)+2\kappa(\nabla_{\bf x}\cdot\nabla_{\bf X})\theta_0
-({\bf V}\cdot\nabla_{\bf X})\theta_0
-({\bf v}_0\cdot\nabla_{\bf x})\theta_0=0.\eqn{22.3}$$
В виду представлений (16), соотношений (12) и (14) при $n=1$ и линейности (22),
эти уравнения имеют решения следующей структуры:
$$\lbd\bom_1\rbd_v=\bxi_1^\omega+\sum_{k=1}^2\left({\bf S}^{v,\omega}_k\lad{\bf v}^k_1\rad
+{\bf S}^{h,\omega}_k\lad{\bf h}^k_1\rad+\sum_{m=1}^2\left(
{\bf G}^{v,\omega}_{m,k}{\partial\lad{\bf v}^k_0\rad\over\partial X_m}
+{\bf G}^{h,\omega}_{m,k}{\partial\lad{\bf h}^k_0\rad\over\partial X_m}\right.\right.$$
$$\left.\left.\phantom{|^|\over|_|}\!\!\!
+{\bf Q}^{vv,\omega}_{m,k}\lad{\bf v}^k_0\rad\lad{\bf v}^m_0\rad
+{\bf Q}^{vh,\omega}_{m,k}\lad{\bf v}^k_0\rad\lad{\bf h}^m_0\rad
+{\bf Q}^{hh,\omega}_{m,k}\lad{\bf h}^k_0\rad\lad{\bf h}^m_0\rad\right)\right),\eqn{23.1}$$
$$\lbd{\bf v}_1\rbd_h=\bxi_1^v+\sum_{k=1}^2\left({\bf S}^{v,v}_k\lad{\bf v}^k_1\rad
+{\bf S}^{h,v}_k\lad{\bf h}^k_1\rad+\sum_{m=1}^2\left(
{\bf G}^{v,v}_{m,k}{\partial\lad{\bf v}^k_0\rad\over\partial X_m}
+{\bf G}^{h,v}_{m,k}{\partial\lad{\bf h}^k_0\rad\over\partial X_m}\right.\right.$$
$$\left.\left.\phantom{|^|\over|_|}\!\!\!
+{\bf Q}^{vv,v}_{m,k}\lad{\bf v}^k_0\rad\lad{\bf v}^m_0\rad
+{\bf Q}^{vh,v}_{m,k}\lad{\bf v}^k_0\rad\lad{\bf h}^m_0\rad
+{\bf Q}^{hh,v}_{m,k}\lad{\bf h}^k_0\rad\lad{\bf h}^m_0\rad\right)\right),\eqn{23.2}$$
$$\lbd{\bf h}_1\rbd_h=\bxi_1^h+\sum_{k=1}^2\left({\bf S}^{v,h}_k\lad{\bf v}^k_1\rad
+{\bf S}^{h,h}_k\lad{\bf h}^k_1\rad+\sum_{m=1}^2\left(
{\bf G}^{v,h}_{m,k}{\partial\lad{\bf v}^k_0\rad\over\partial X_m}
+{\bf G}^{h,h}_{m,k}{\partial\lad{\bf h}^k_0\rad\over\partial X_m}\right.\right.$$
$$\left.\left.\phantom{|^|\over|_|}\!\!\!
+{\bf Q}^{vv,h}_{m,k}\lad{\bf v}^k_0\rad\lad{\bf v}^m_0\rad
+{\bf Q}^{vh,h}_{m,k}\lad{\bf v}^k_0\rad\lad{\bf h}^m_0\rad
+{\bf Q}^{hh,h}_{m,k}\lad{\bf h}^k_0\rad\lad{\bf h}^m_0\rad\right)\right),\eqn{23.3}$$
$$\theta_1=\xi_1^\theta+\sum_{k=1}^2\left(S^{v,\theta}_k\lad{\bf v}^k_1\rad
+S^{h,\theta}_k\lad{\bf h}^k_1\rad+\sum_{m=1}^2\left(
G^{v,\theta}_{m,k}{\partial\lad{\bf v}^k_0\rad\over\partial X_m}
+G^{h,\theta}_{m,k}{\partial\lad{\bf h}^k_0\rad\over\partial X_m}\right.\right.$$
$$\left.\left.\phantom{|^|\over|_|}\!\!\!
+Q^{vv,\theta}_{m,k}\lad{\bf v}^k_0\rad\lad{\bf v}^m_0\rad
+Q^{vh,\theta}_{m,k}\lad{\bf v}^k_0\rad\lad{\bf h}^m_0\rad
+Q^{hh,\theta}_{m,k}\lad{\bf h}^k_0\rad\lad{\bf h}^m_0\rad\right)\right),\eqn{23.4}$$
где функции ${\bf G}^{\cdot,\cdot}_{m,k}$ являются решениями {\it вспомогательных задач второго типа}:
\pagebreak
$${\cal L}^\omega({\bf G}^{v,\omega}_{m,k},{\bf G}^{v,v}_{m,k},{\bf G}^{v,h}_{m,k},G^{v,\theta}_{m,k})=
-2\nu{\partial{\bf S}^{v,\omega}_k\over\partial x_m}
-\epsilon_{m,k,3}{\partial{\bf V}\over\partial x_3}
-{\bf e}_m\times\left({\bf V}\times{\bf S}_k^{v,\omega\phantom{\theta}}\right.$$
$$\left.+({\bf S}^{v,v}_k+{\bf e}_k)\times\bOm
-{\bf H}\times(\nabla_{\bf x}\times{\bf S}^{v,h}_k)
-{\bf S}^{v,h}_k\times(\nabla_{\bf x}\times{\bf H})
+\beta S^{v,\theta}_k{\bf e}_3\right)$$
$$-\nabla_{\bf x}\times({\bf H}\times({\bf e}_m\times{\bf S}^{v,h}_k)),\eqn{24.1}$$
$$\nabla_{\bf x}\cdot{\bf G}^{v,\omega}_{m,k}=-({\bf S}^{v,\omega}_k)^m,\eqn{24.2}$$
$$\nabla_{\bf x}\times{\bf G}^{v,v}_{m,k}={\bf G}^{v,\omega}_{m,k}
-{\bf e}_m\times{\bf S}^{v,v}_k,\eqn{24.3}$$
$$\nabla_{\bf x}\cdot{\bf G}^{v,v}_{m,k}=-({\bf S}^{v,v}_k)^m,\eqn{24.4}$$
$${\cal L}^h({\bf G}^{v,v}_{m,k},{\bf G}^{v,h}_{m,k})=
-2\eta{\partial{\bf S}^{v,h}_k\over\partial x_m}
-{\bf e}_m\times\left({\bf V}\times{\bf S}^{v,h}_k
+({\bf S}^{v,v}_k+{\bf e}_k)\times{\bf H}\right),\eqn{24.5}$$
$$\nabla_{\bf x}\cdot{\bf G}^{v,h}_{m,k}=-({\bf S}^{v,h}_k)^m,\eqn{24.6}$$
$${\cal L}^\theta({\bf G}^{v,v}_{m,k},G^{v,\theta}_{m,k})=
-2\kappa{\partial S^{v,\theta}_k\over\partial x_m}+{\bf V}^mS^{v,\theta}_k\eqn{24.7}$$
(здесь $\epsilon_{m,k,j}$ -- стандартный единичный антисимметричный тензор);
$${\cal L}^\omega({\bf G}^{h,\omega}_{m,k},{\bf G}^{h,v}_{m,k},{\bf G}^{h,h}_{m,k},G^{h,\theta}_{m,k})=
-2\nu{\partial{\bf S}^{h,\omega}_k\over\partial x_m}
-{\bf e}_m\times\left({\bf V}\times{\bf S}^{h,\omega}_k\right.$$
$$\left.+{\bf S}^{h,v}_k\times\bOm
-{\bf H}\times(\nabla_{\bf x}\times{\bf S}^{h,h}_k)
-({\bf S}^{h,h}_k+{\bf e}_k)\times(\nabla_{\bf x}\times{\bf H})
+\beta S^{h,\theta}_k{\bf e}_3\right)$$
$$-\nabla_{\bf x}\times({\bf H}\times({\bf e}_m\times{\bf S}^{h,h}_k)),\eqn{25.1}$$
$$\nabla_{\bf x}\cdot{\bf G}^{h,\omega}_{m,k}=-({\bf S}^{h,\omega}_k)^m,\eqn{25.2}$$
$$\nabla_{\bf x}\times{\bf G}^{h,v}_{m,k}={\bf G}^{h,\omega}_{m,k}
-{\bf e}_m\times{\bf S}^{h,v}_k,\eqn{25.3}$$
$$\nabla_{\bf x}\cdot{\bf G}^{h,v}_{m,k}=-({\bf S}^{h,v}_k)^m,\eqn{25.4}$$
$${\cal L}^h({\bf G}^{h,v}_{m,k},{\bf G}^{h,h}_{m,k})=
-2\eta{\partial{\bf S}^{h,h}_k\over\partial x_m}
-{\bf e}_m\times\left({\bf V}\times({\bf S}^{h,h}_k+{\bf e}_k)
+{\bf S}^{h,v}_k\times{\bf H}\right),\eqn{25.5}$$
$$\nabla_{\bf x}\cdot{\bf G}^{h,h}_{m,k}=-({\bf S}^{h,h}_k)^m,\eqn{25.6}$$
$${\cal L}^\theta({\bf G}^{h,v}_{m,k},G^{h,\theta}_{m,k})=
-2\kappa{\partial S^{h,\theta}_k\over\partial x_m}
+{\bf V}^mS^{h,\theta}_k;\eqn{25.7}$$
функции ${\bf Q}^{\cdot,\cdot\cdot}_{m,k}$ -- решениями {\it вспомогательных задач третьего типа}:
$$\left.{\cal L}^\omega({\bf Q}^{vv,\omega}_{m,k},{\bf Q}^{vv,v}_{m,k},{\bf Q}^{vv,h}_{m,k},Q^{vv,\theta}_{m,k})=
\rho_{m,k}\nabla_{\bf x}\times\right(
-({\bf S}^{v,v}_k+{\bf e}_k)\times{\bf S}^{v,\omega}_m$$
$$\left.+{\bf S}^{v,h}_k\times(\nabla_{\bf x}\times{\bf S}^{v,h}_m)
-({\bf S}^{v,v}_m+{\bf e}_m)\times{\bf S}^{v,\omega}_k
+{\bf S}^{v,h}_m\times(\nabla_{\bf x}\times{\bf S}^{v,h}_k)\right),\eqn{26.1}$$
$$\nabla_{\bf x}\times{\bf Q}^{vv,v}_{m,k}={\bf Q}^{vv,\omega}_{m,k},\eqn{26.2}$$
$${\cal L}^h({\bf Q}^{vv,v}_{m,k},{\bf Q}^{vv,h}_{m,k})=-\rho_{m,k}\nabla_{\bf x}\times
\left(({\bf S}^{v,v}_k+{\bf e}_k)\times{\bf S}^{v,h}_m
+({\bf S}^{v,v}_m+{\bf e}_m)\times{\bf S}^{v,h}_k\right),\eqn{26.3}$$
$${\cal L}^\theta({\bf Q}^{vv,v}_{m,k},Q^{vv,\theta}_{m,k})=\rho_{m,k}\left(
(({\bf S}^{v,v}_k+{\bf e}_k)\cdot\nabla_{\bf x})S^{v,\theta}_m
+(({\bf S}^{v,v}_m+{\bf e}_m)\cdot\nabla_{\bf x})S^{v,\theta}_k\right);\eqn{26.4}$$
$$\left.{\cal L}^\omega({\bf Q}^{vh,\omega}_{m,k},{\bf Q}^{vh,v}_{m,k},{\bf Q}^{vh,h}_{m,k},Q^{vh,\theta}_{m,k})=
-\nabla_{\bf x}\times\right(({\bf S}^{v,v}_k+{\bf e}_k)\times{\bf S}^{h,\omega}_m
+{\bf S}^{h,v}_m\times{\bf S}^{v,\omega}_k$$
$$\left.-{\bf S}^{v,h}_k\times(\nabla_{\bf x}\times{\bf S}^{h,h}_m)
-({\bf S}^{h,h}_m+{\bf e}_m)\times(\nabla_{\bf x}\times{\bf S}^{v,h}_k)\right),\eqn{27.1}$$
$$\nabla_{\bf x}\times{\bf Q}^{vh,v}_{m,k}={\bf Q}^{vh,\omega}_{m,k},\eqn{27.2}$$
$${\cal L}^h({\bf Q}^{vh,v}_{m,k},{\bf Q}^{vh,h}_{m,k})=-\nabla_{\bf x}\times
\left(({\bf S}^{v,v}_k+{\bf e}_k)\times({\bf S}^{h,h}_m+{\bf e}_m)+
{\bf S}^{h,v}_m\times{\bf S}^{v,h}_k\right),\eqn{27.3}$$
$${\cal L}^\theta({\bf Q}^{vh,v}_{m,k},Q^{vh,\theta}_{m,k})=
(({\bf S}^{v,v}_k+{\bf e}_k)\cdot\nabla_{\bf x})S^{h,\theta}_m
+({\bf S}^{h,v}_m\cdot\nabla_{\bf x})S^{v,\theta}_k;\eqn{27.4}$$
$${\cal L}^\omega({\bf Q}^{hh,\omega}_{m,k},{\bf Q}^{hh,v}_{m,k},{\bf Q}^{hh,h}_{m,k},Q^{hh,\theta}_{m,k})=
\rho_{m,k}\nabla_{\bf x}\times\left(-{\bf S}^{h,v}_k\times{\bf S}^{h,\omega}_m\right.$$
$$\left.+({\bf S}^{h,h}_k+{\bf e}_k)\times(\nabla_{\bf x}\times{\bf S}^{h,h}_m)
-{\bf S}^{h,v}_m\times{\bf S}^{h,\omega}_k
+({\bf S}^{h,h}_m+{\bf e}_m)\times(\nabla_{\bf x}\times{\bf S}^{h,h}_k)\right),\eqn{28.1}$$
$$\nabla_{\bf x}\times{\bf Q}^{hh,v}_{m,k}={\bf Q}^{hh,\omega}_{m,k},\eqn{28.2}$$
$${\cal L}^h({\bf Q}^{hh,v}_{m,k},{\bf Q}^{hh,h}_{m,k})=-\rho_{m,k}\nabla_{\bf x}\times
\left({\bf S}^{h,v}_k\times({\bf S}^{h,h}_m+{\bf e}_m)
+{\bf S}^{h,v}_m\times({\bf S}^{h,h}_k+{\bf e}_k)\right),\eqn{28.3}$$
$${\cal L}^\theta({\bf Q}^{hh,v}_{m,k},Q^{hh,\theta}_{m,k})=\rho_{m,k}\left(
({\bf S}^{h,v}_k\cdot\nabla_{\bf x})S^{h,\theta}_m
+({\bf S}^{h,v}_m\cdot\nabla_{\bf x})S^{h,\theta}_k\right);\eqn{28.4}$$
$$\nabla_{\bf x}\cdot{\bf Q}^{\cdot\cdot,\cdot}_{m,k}=0\eqn{29}$$
(задачи (26) и (28) определены для $m\le k$;
$\rho_{m,k}=1$ при $m<k$, и $\rho_{m,k}=1/2$ при $m=k$;
при $m>k$ считаем ${\bf Q}^{vv,\cdot}_{m,k}={\bf Q}^{hh,\cdot}_{m,k}=0$);
а функции $\bxi_1$ -- решениями задач
$${\cal L}^\omega(\bxi^\omega_1,\bxi^v_1,\bxi^h_1,\xi^\theta_1)
+2\nu(\nabla_{\bf x}\cdot\nabla_{\bf X})\bxi^\omega_0
-\nabla_{\bf x}\times({\bf H}\times(\nabla_{\bf X}\times\bxi^h_0))$$
$$+\nabla_{\bf X}\times\left({\bf V}\times\bxi^\omega_0+\bxi^v_0\times\bOm
-{\bf H}\times(\nabla_{\bf x}\times\bxi^h_0)-\bxi^h_0\times(\nabla_{\bf x}\times{\bf H})\right)$$
$$+\nabla_{\bf x}\times\left(
{\bf v}_0\times\bxi^\omega_0+\bxi^v_0\times(\bom_0-\bxi^\omega_0)
-{\bf h}_0\times(\nabla_{\bf x}\times\bxi^h_0)\right.$$
$$\left.-\bxi^h_0\times(\nabla_{\bf x}\times({\bf h}_0-\bxi^h_0))\right)
+\beta\nabla_{\bf X}\xi^\theta_0\times{\bf e}_3=0,\eqn{30.1}$$
$$\nabla_{\bf x}\cdot\bxi_1^\omega+\nabla_{\bf X}\cdot\bxi_0^\omega=0,\eqn{30.2}$$
$$\nabla_{\bf x}\times\bxi^v_1=\bxi^\omega_1-\nabla_{\bf X}\times\bxi^v_0,\eqn{30.3}$$
$$\nabla_{\bf x}\cdot\bxi_1^v+\nabla_{\bf X}\cdot\bxi_0^v=0,\eqn{30.4}$$
$${\cal L}^h(\bxi^v_1,\bxi^h_1)+2\eta(\nabla_{\bf x}\cdot\nabla_{\bf X})\bxi^h_1
+\nabla_{\bf X}\times(\bxi^v_0\times{\bf H}+{\bf V}\times\bxi^h_0)$$
$$+\nabla_{\bf x}\times({\bf v}_0\times\bxi^h_0
+\bxi^v_0\times({\bf h}_0-\bxi^h_0))=0,\eqn{30.5}$$
$$\nabla_{\bf x}\cdot\bxi_1^h+\nabla_{\bf X}\cdot\bxi_0^h=0,\eqn{30.6}$$
$${\cal L}^\theta(\bxi^v_1,\xi^\theta_1)
+2\kappa(\nabla_{\bf x}\cdot\nabla_{\bf X})\xi^\theta_0
-({\bf V}\cdot\nabla_{\bf X})\xi^\theta_0
-({\bf v}_0\cdot\nabla_{\bf x})\xi^\theta_0
-(\bxi^v_0\cdot\nabla_{\bf x})(\theta_0-\xi^\theta_0)=0.\eqn{30.7}$$
Уравнения (26.1), (27.1) и (28.1) эквивалентны уравнениям

\pagebreak
$$\left.{\cal L}^v({\bf Q}^{vv,v}_{m,k},{\bf Q}^{vv,h}_{m,k},Q^{vv,\theta}_{m,k},Q^{vv,p}_{m,k})=
\rho_{m,k}\right(-({\bf S}^{v,v}_k+{\bf e}_k)\times{\bf S}^{v,\omega}_m$$
$$\left.+{\bf S}^{v,h}_k\times(\nabla_{\bf x}\times{\bf S}^{v,h}_m)
-({\bf S}^{v,v}_m+{\bf e}_m)\times{\bf S}^{v,\omega}_k
+{\bf S}^{v,h}_m\times(\nabla_{\bf x}\times{\bf S}^{v,h}_k)\right),\eqn{26.1'}$$
$${\cal L}^v({\bf Q}^{vh,v}_{m,k},{\bf Q}^{vh,h}_{m,k},Q^{vh,\theta}_{m,k},Q^{vh,p}_{m,k})=
-({\bf S}^{v,v}_k+{\bf e}_k)\times{\bf S}^{h,\omega}_m
-{\bf S}^{h,v}_m\times{\bf S}^{v,\omega}_k$$
$$+{\bf S}^{v,h}_k\times(\nabla_{\bf x}\times{\bf S}^{h,h}_m)
+({\bf S}^{h,h}_m+{\bf e}_m)\times(\nabla_{\bf x}\times{\bf S}^{v,h}_k),\eqn{27.1'}$$
$${\cal L}^v({\bf Q}^{hh,v}_{m,k},{\bf Q}^{hh,h}_{m,k},Q^{hh,\theta}_{m,k},Q^{hh,p}_{m,k})=
\rho_{m,k}\left(-{\bf S}^{h,v}_k\times{\bf S}^{h,\omega}_m\right.$$
$$\left.+({\bf S}^{h,h}_k+{\bf e}_k)\times(\nabla_{\bf x}\times{\bf S}^{h,h}_m)
-{\bf S}^{h,v}_m\times{\bf S}^{h,\omega}_k
+({\bf S}^{h,h}_m+{\bf e}_m)\times(\nabla_{\bf x}\times{\bf S}^{h,h}_k)\right),\eqn{28.1'}$$
где $\la\nabla_{\bf x}Q^{\cdot\cdot,p}_{m,k}\ra_h=0$.
Условия (24.3), (25.3), (26.2), (27.2), (28.2) и (30.3) получаются
при подстановке (23.1) и (23.2) в (13) с учетом (14). Аналогично,
(24.2), (25.2), $\nabla_{\bf x}\cdot{\bf Q}^{\cdot\cdot,\omega}_{m,k}=0$ и
(30.2) -- следствие (12.5) при \hbox{$n=1$} (а также могут быть получены взятием дивергенции
уравнений (24.3), (25.3), (26.2), (27.2), (28.2) и (30.3)); (24.4), (25.4),
$\nabla_{\bf x}\cdot{\bf Q}^{\cdot\cdot,v}_{m,k}=0$ и (30.4) -- следствие
(12.2), а (24.6), (25.6), $\nabla_{\bf x}\cdot{\bf Q}^{\cdot\cdot,h}_{m,k}=0$
и (30.6) -- следствие (12.3).

Из дивергенций уравнений (26.3), (27.3) и (28.3) находим, что поля
${\bf Q}^{\cdot\cdot,h}_{m,k}$ соленоидальны при $t>0$, если и только если
они соленоидальны при $t=0$. Взяв дивергенцию уравнений (24.1) и (25.1)
и используя (18.1) и (19.1), соответственно, получаем
$$\left(-{\partial\over\partial t}+\eta\Delta_{\bf x}\right)
\left(\nabla_{\bf x}\cdot{\bf G}^{\cdot,\omega}_{m,k}+({\bf S}^{\cdot,\omega}_k)^m\right)=0,$$
откуда (24.2) и (25.2) выполнены при $t>0$ тогда и только тогда, когда
они выполнены при $t=0$. Аналогичным образом, из дивергенций уравнений (24.5),
(25.5), (30.1) и (30.5) совместно с уравнениями (18.3), (19.3) и (20.1),
соответствен\-но, находим, что (24.6), (25.6), (30.2) и (30.6) выполнены при $t>0$,
если и только если они выполнены при $t=0$. Условия (24.2), (25.2), (30.2)
и (29) совместно гарантируют выполнение (12.5) при $n=1$, а (24.6), (25.6),
(30.6) и (29) -- (12.3) при $n=1$. Поэтому в качестве начальных условий
для задач (24)-(30) выбираем глобально ограниченные вместе с производными
гладкие поля, удовлетворяющие (24.2), (24.6), (25.2), (25.6), (29), (30.2)
и (30.6), а также перечисленным ниже условиям.

Мы полагаем, что каждое слагаемое в суммах по $k$ и $m$ в (23) имеет нулевое
среднее соответствующих компонент:
$$\lad{\bf G}^{\cdot,\omega}_{m,k}\rad_v=
\lad{\bf Q}^{\cdot\cdot,\omega}_{m,k}\rad_v=0,\quad
\lad{\bf G}^{\cdot,h}_{m,k}\rad_h=\lad{\bf Q}^{\cdot\cdot,h}_{m,k}\rad_h=0.$$
Интегрируя (24.5) и (25.5) по быстрому времени от 0 до $t$ и
усредняя результат по быстрым переменным
с учетом равенств (7.2) и $\lad{\bf G}^{\cdot,\omega}_{m,k}\rad_v=0$, находим
$$\la{\bf G}^{v,h}_{m,k}\ra_h|_{t=0}
=\ladb{\bf e}_m\times\int_0^t\left({\bf V}\times{\bf S}^{v,h}_k
+({\bf S}^{v,v}_k+{\bf e}_k)\times{\bf H}\right)\,dt\radb_h,\eqn{31.1}$$
$$\la{\bf G}^{h,h}_{m,k}\ra_h|_{t=0}
=\ladb{\bf e}_m\times\int_0^t\left({\bf V}\times({\bf S}^{h,h}_k+{\bf e}_k)
+{\bf S}^{h,v}_k\times{\bf H}\right)\,dt\radb_h.\eqn{31.2}$$
Магнитный $\alpha-$эффект несущественен, если существуют средние в правых
частях (31). Очевидно, пространственные средние вертикальных компонент
подынтег\-ральных выражений в (31) равны 0, и это условие
выполнено, если исходное конвективное МГД состояние центрально-симметрично
или симметрично относи\-тельно вертикальной оси, т.к. тогда исходные
поля $\bf V$ и $\bf H$, и поля ${\bf S}^{\cdot,\cdot}_k$ имеют противоположные
типы симметрий. Это более сильное условие, чем условия (8.2), необходимые для
существования решений системы уравнений (9), которые в приложении к системам
(24) и (25) в силу равенств $\lad{\bf V}\rad_h=\lad{\bf H}\rad_h=0$ имеют вид
$$\lad{\bf V}\times{\bf S}^{v,h}_k+{\bf S}^{v,v}_k\times{\bf H}\rad_v=0;\qquad
\lad{\bf V}\times{\bf S}^{h,h}_k+{\bf S}^{h,v}_k\times{\bf H}\rad_v=0,\eqn{32}$$
соответственно. Отметим, что мы не требуем равенства нулю горизонтальных
компонент $\alpha-$тензоров в уравнении магнитного поля.

Усредняя (24.1), (25.1), (26.1), (26.3), (27.1), (27.3), (28.1) и (28.3)
по быстрым пространственным переменным с учетом (7.1) и (7.2), а также равенств\break
$\bOm=\nabla_{\bf x}\times\bf V$,
${\bf S}^{\cdot,\omega}_k=\nabla_{\bf x}\times{\bf S}^{\cdot,v}_k$ и
$\nabla_{\bf x}\cdot{\bf V}=\nabla_{\bf x}\cdot{\bf S}^{\cdot,v}_k=0$
при усреднении (24.1) и (25.1), находим, что
$\la{\bf G}^{\cdot,\omega}_{m,k}\ra_v$,
$\la{\bf Q}^{\cdot\cdot,\omega}_{m,k}\ra_v$ и
$\la{\bf Q}^{\cdot\cdot,h}_{m,k}\ra_h$ не зависят от времени,
условия (8.1), необходимые для существования решений системы уравнений (9),
для систем (24) и (25) выполнены, и должны иметь место равенства
$$\la{\bf G}^{\cdot,\omega}_{m,k}\ra_v|_{t=0}=
\la{\bf Q}^{\cdot\cdot,\omega}_{m,k}\ra_v|_{t=0}=0,\quad
\la{\bf Q}^{\cdot\cdot,h}_{m,k}\ra_h|_{t=0}=0.\eqn{33}$$

Интегрируя (30.1) по быстрому времени и усредняя результат
по быстрым пространственным переменным с учетом (7.1), а также равенств
$\bOm=\nabla_{\bf x}\times\bf V$, $\bxi^\omega_0=\nabla_{\bf x}\times\bxi^v_0$
и $\nabla_{\bf x}\cdot{\bf V}=\nabla_{\bf x}\cdot\bxi^v_0=0$, находим, что
$\la\bxi_1^\omega\ra_v$ не зависит от быстрого времени. Поэтому из
$\lad\bxi_1^\omega\rad_v=0$ (см. (23.1)) следует $\la\bxi_1^\omega\ra_v|_{t=0}=0$,
и поскольку пространственное среднее каждой суммы по $k$ в (23.1) равно 0,
$$\lad\bom_1\rad_v|_{T=0}=\la\bom_1\ra_v|_{t=0}.$$
Аналогично, из (23.3)
$$\lad{\bf h}_1\rad_h|_{T=0}=\la{\bf h}_1\ra_h|_{t=0}-\la\bxi_1^h\ra_h|_{t=0}$$
$$-\sum_{k=1}^2\sum_{m=1}^2\left(
\la{\bf G}^{v,h}_{m,k}\ra_h|_{t=0}{\partial\lad{\bf v}^k_0\rad\over\partial X_m}
+\la{\bf G}^{h,h}_{m,k}\ra_h|_{t=0}{\partial\lad{\bf h}^k_0\rad\over\partial X_m}\right),$$
а из (30.5)
$$\la\bxi_1^h\ra_h|_{t=0}=\ladb\int_0^t\nabla_{\bf X}\times\left(
{\bf V}\times\bxi_0^h+\bxi_0^v\times{\bf H}\right)\,dt\radb_h.$$
Отсюда, зная $\bom_1|_{t=0}$ и ${\bf h}_1|_{t=0}$, находим
$\lad\bom_1\rad|_{T=0}$ и $\lad{\bf h}_1\rad|_{T=0}$, а затем из (23.1)
и (23.3), подставив туда $\lbd\bom_1\rbd_v=\bom_1-\lad\bom_1\rad_v$ и
$\lbd{\bf h}_1\rbd_h={\bf h}_1-\lad{\bf h}_1\rad_h$,
находим начальные условия для задачи (30). Изменение начальных условий для
${\bf G}^{\cdot,\cdot}_{m,k}$ и ${\bf Q}^{\cdot\cdot,\cdot}_{m,k}$
в рассматриваемом классе начальных условий компенсируются соответству\-ющим
изменением начальных условий для $\bxi^\cdot_1$. Вызванные этим
изменения полей ${\bf G}^{\cdot,\cdot}_{m,k}$ и ${\bf Q}^{\cdot\cdot,\cdot}_{m,k}$
экспоненциально затухают во времени (т.к. они являются решениями задачи (10)
с нулевыми пространственными средними).

Поскольку функции $\bxi_0$ и их производные экспоненциально убывают
во време\-ни (см. часть 5), то же верно для правых частей (30). Как показал,
используя принцип Дюамеля, Желиговский [2006], тогда
из предположения о линейной устойчивости исходного
конвективного МГД состояния по отношению к коротко\-масштабным возмущениям
следует, что функции $\bxi_1$ и их производные также экспоненциально
убывают во времени. Таким образом, $\bxi_1$ имеют смысл затухаю\-щих
переходных процессов. Аналогично доказывается, что экспоненциально убы\-вают
во времени изменения ${\bf G}^{\cdot,\cdot}_{m,k}$ и
${\bf Q}^{\cdot\cdot,\cdot}_{m,k}$, вызванные возможной неоднозначностью
${\bf S}^{\cdot,\cdot}_k$. Как следует из выведенных ниже выражений для
коэффициентов в новых членах, появляющихся в усредненных уравнениях,
экспоненциально затухающие изменения этих полей не изменяют величины
коэффициентов. Таким образом, выбор начальных условий не влияет
на вид уравнений для средних полей.

Если исходное МГД состояние имеет симметрию относительно центра или
вертикальной оси, то поля ${\bf S}^{\cdot,\cdot}_k$ имеют противоположный тип
симметрии, а тогда правые части уравнений (24)-(28) имеют тот же тип симметрии,
что и исходное конвективное МГД состояние. Удобно потребовать, чтобы при
наличии у исходно\-го состояния симметрии одного из указанных типов
${\bf G}^{\cdot,\cdot}_{m,k}|_{t=0}$ и ${\bf Q}^{\cdot\cdot,\cdot}_{m,k}|_{t=0}$
также обладали ею. Поскольку пространства симметричных и антисимметричных
состо\-яний тогда инвариантны для оператора ${\cal L}=({\cal L}^\omega,{\cal L}^h,
{\cal L}^\theta)$, то поля ${\bf G}^{\cdot\cdot,\cdot}_{m,k}$ и
${\bf Q}^{\cdot\cdot,\cdot}_{m,k}$ имеют тогда тот же тип симметрии,
что и у исходного состояния. В этом случае в силу антисимметрии
${\bf S}^{\cdot,\cdot}_k$ вертикальные компоненты подынтегральных выражений
в (31) равны 0, и (31)-(33) автоматически выполнены.

В случае, если исходное МГД
состояние стационарно или периодично по (быст\-рому) времени,
для удобства вычисления прост\-ранственно-временн\'ых средних
естественно потребовать, чтобы функции ${\bf G}^{\cdot,\cdot}_{m,k}$ и
${\bf Q}^{\cdot\cdot,\cdot}_{m,k}$ были, соответственно,
стационарными или периодичными по времени решениями задач (19)-(24).
Су\-ществование таких стационарных решений рассмотрено Желиговским [2003];
периодический по времени случай рассматривается аналогично (см. Zheligovsky
и Podvigina [2003]).

\mi{\bf 7. Уравнения порядка $\varepsilon^2$.}
Уравнения, полученные из членов рядов (5.1)-(5.3) порядка $\varepsilon^2$,
имеют вид
$${\cal L}^\omega(\bom_2,{\bf v}_2,{\bf h}_2,\theta_2)
-{\partial\bom_0\over\partial T}
+\nu(2(\nabla_{\bf x}\cdot\nabla_{\bf X})\lbd\bom_1\rbd_v
+\Delta_{\bf X}\bom_0)$$
$$+\nabla_{\bf X}\times({\bf V}\times\bom_1+{\bf v}_1\times\bOm
-{\bf H}\times(\nabla_{\bf X}\times{\bf h}_0+
\nabla_{\bf x}\times{\bf h}_1)$$
$$-{\bf h}_1\times(\nabla_{\bf x}\times{\bf H})
+{\bf v}_0\times\bom_0-{\bf h}_0\times(\nabla_{\bf x}\times{\bf h}_0))$$
$$+\nabla_{\bf x}\times(-{\bf H}\times(\nabla_{\bf X}\times{\bf h}_1)
+{\bf v}_1\times\bom_0+{\bf v}_0\times\bom_1
-{\bf h}_1\times(\nabla_{\bf x}\times{\bf h}_0)$$
$$-{\bf h}_0\times(\nabla_{\bf x}\times{\bf h}_1+\nabla_{\bf X}\times{\bf h}_0))
+\beta\nabla_{\bf X}\theta_1\times{\bf e}_3=0;\eqn{34.1}$$
$${\cal L}^h({\bf v}_2,{\bf h}_2)-{\partial{\bf h}_0\over\partial T}
+\eta(2(\nabla_{\bf x}\cdot\nabla_{\bf X})\lbd{\bf h}_1\rbd_h+\Delta_{\bf X}{\bf h}_0)$$
$$+\nabla_{\bf X}\times({\bf v}_1\times{\bf H}+{\bf V}\times{\bf h}_1+{\bf v}_0\times{\bf h}_0)
+\nabla_{\bf x}\times({\bf v}_1\times{\bf h}_0+{\bf v}_0\times{\bf h}_1)=0;\eqn{34.2}$$
\pagebreak
$${\cal L}^\theta({\bf v}_2,\theta_2)-{\partial\theta_0\over\partial T}
+\kappa(2(\nabla_{\bf x}\cdot\nabla_{\bf X})\theta_1+\Delta_{\bf X}\theta_0)$$
$$-({\bf V}\cdot\nabla_{\bf X})\theta_1
-({\bf v}_1\cdot\nabla_{\bf x})\theta_0
-({\bf v}_0\cdot\nabla_{\bf x})\theta_1
-({\bf v}_0\cdot\nabla_{\bf X})\theta_0=0.\eqn{34.3}$$

Усреднение вертикальной компоненты (34.1) по быстрым про\-странственным
переменным с учетом (7.1), (12.1)-(12.3), (13) при $n=0,1$, (14) при $n=0$,
краевых условий для $\bf V$, $\bf H$, ${\bf v}_i$ и ${\bf h}_i$
и соленоидальности ${\bf v}_0$ и ${\bf h}_0$ дает
$$-{\partial\la\bom_2\ra_v\over\partial t}+\nabla_{\bf X}\times\la
{\bf V}\times(\nabla_{\bf X}\times{\bf v}_0)
-{\bf V}\nabla_{\bf X}\cdot\lbd{\bf v}_0\rbd_h$$
$$-{\bf H}\times(\nabla_{\bf X}\times{\bf h}_0)
+{\bf H}\nabla_{\bf X}\cdot\lbd{\bf h}_0\rbd_h\ra_h=0.$$
Проинтегрировав это уравнение по быстрому времени от 0 до $t$,
подставив пред\-ставления (16.2) и (16.3) и усреднив результат по быстрому
времени, с учетом экспоненциального убывания $\bxi_0^v$ и $\bxi_0^h$
во времени, получим, что $\la\bom_2\ra_v|_{t=0}$ конечно, если существуют средние
$\lad\int_0^t\bal^{\cdot}_{m,k}\,dt\rad_h$ для $k,m=1,2,\ k\ne m$,
$\lad\int_0^t(\bal^v_{1,1}-\bal^v_{2,2})\,dt\rad_h$ и
$\lad\int_0^t(\bal^h_{1,1}-\bal^h_{2,2})\,dt\rad_h$.
Здесь обозначено
$$\bal^v_{m,k}={\bf V}\times({\bf e}_m\times({\bf S}^{v,v}_k+{\bf e}_k))
-{\bf H}\times({\bf e}_m\times{\bf S}^{v,h}_k)
-{\bf V}({\bf S}^{v,v}_k)^m+{\bf H}({\bf S}^{v,h}_k)^m,\eqn{35.1}$$
$$\bal^h_{m,k}={\bf V}\times({\bf e}_m\times{\bf S}^{h,v}_k)
-{\bf H}\times({\bf e}_m\times({\bf S}^{h,h}_k+{\bf e}_k))
-{\bf V}({\bf S}^{h,v}_k)^m+{\bf H}({\bf S}^{h,h}_k)^m.\eqn{35.2}$$
Кинематический $\alpha-$эффект несущественен, если эти средние существуют.
Оче\-видно, пространственные средние горизонтальных компонент
(35) равны 0, и тем самым это условие выполнено,
если исходное конвективное МГД состояние центрально-симметрично
или симметрично относительно вертикальной оси, т.к. тогда исходные
поля $\bf V$ и $\bf H$ имеют тип симметрии, противоположный типу симметрии
${\bf S}^{h,h}_k$ и ${\bf S}^{h,v}_k$.
Это более сильное условие, чем условия (8.1), необходимые для
существования решений системы уравнений (9), которые в приложении к (34)
имеют вид
$$\lad\bal^{\cdot}_{m,k}\rad_h=0\mbox{ для }k,m=1,2,\ k\ne m\mbox{ и }
\lad\bal^v_{1,1}-\bal^v_{2,2}\rad_h=\lad\bal^h_{1,1}-\bal^h_{2,2}\rad_h=0.$$
Отметим, что мы не требуем равенства нулю вертикальных
компонент кинемати\-ческого $\alpha-$тензора в уравнении для завихренности.

Усреднение горизонтальных компонент (34.2) по быстрым про\-странственным
переменным с учетом (7.2), (16.2), (16.3), (32) и
краевых условий для $\bf V$, $\bf H$, ${\bf v}_i$ и ${\bf h}_i$ дает
$$-{\partial\over\partial T}\lad{\bf h}_0\rad_h+\eta\Delta_{\bf X}\lad{\bf h}_0\rad_h
+\nabla_{\bf X}\times\left(\sum_{m=1}^2\sum_{k=1}^2
\left({\bf D}^{v,h}_{m,k}{\partial\lad{\bf v}_0^k\rad\over\partial X_m}
+{\bf D}^{h,h}_{m,k}{\partial\lad{\bf h}_0^k\rad\over\partial X_m}\right)\right.$$
$$+\lad{\bf v}_0\rad_h\times\lad{\bf h}_0\rad_h
+\sum_{m=1}^2\sum_{k=1}^2\left(
{\bf A}^{vv,h}_{m,k}\lad{\bf v}^k_0\rad\lad{\bf v}^m_0\rad\right.$$
$$\left.\phantom{\partial\over\partial}
\left.+{\bf A}^{vh,h}_{m,k}\lad{\bf v}^k_0\rad\lad{\bf h}^m_0\rad
+{\bf A}^{hh,h}_{m,k}\lad{\bf h}^k_0\rad\lad{\bf h}^m_0\rad\right)\right)=0.\eqn{36}$$
Здесь $\bf D$ обозначают коэффициенты членов уравнения среднего магнитного
поля, отвечающие т.н. вихревой коррекции магнитной диффузии, $\bf A$ --
квадратичных членов т.н. вихревой коррекции адвекции магнитного поля:
$${\bf D}^{v,h}_{m,k}=\lad{\bf V}\times{\bf G}^{v,h}_{m,k}-{\bf H}\times{\bf G}^{v,v}_{m,k}\rad_v,\eqn{37.1}$$
$${\bf D}^{h,h}_{m,k}=\lad{\bf V}\times{\bf G}^{h,h}_{m,k}-{\bf H}\times{\bf G}^{h,v}_{m,k}\rad_v,\eqn{37.2}$$
$${\bf A}^{vv,h}_{m,k}=\lad{\bf V}\times{\bf Q}^{vv,h}_{m,k}-{\bf H}\times{\bf Q}^{vv,v}_{m,k}
+{\bf S}^{v,v}_k\times{\bf S}^{v,h}_m\rad_v,\eqn{38.1}$$
$${\bf A}^{vh,h}_{m,k}=\lad{\bf V}\times{\bf Q}^{vh,h}_{m,k}-{\bf H}\times{\bf Q}^{vh,v}_{m,k}
+{\bf S}^{v,v}_k\times{\bf S}^{h,h}_m+{\bf S}^{h,v}_m\times{\bf S}^{v,h}_k\rad_v,\eqn{38.2}$$
$${\bf A}^{hh,h}_{m,k}=\lad{\bf V}\times{\bf Q}^{hh,h}_{m,k}-{\bf H}\times{\bf Q}^{hh,v}_{m,k}
+{\bf S}^{h,v}_k\times{\bf S}^{h,h}_m\rad_v.\eqn{38.3}$$

При выполнении указанных условий разрешимости уравнений (34) их решения
можно представить как сумму экспоненциально затухающего во времени слагаемо\-го,
определяемого начальными условиями, и членов, пропорциональных
$${\partial\lad{\bf h}^k_0\rad\over\partial T},
{\partial\lad{\bf v}^k_0\rad\over\partial T},
{\partial^2\lad{\bf v}^k_0\rad\over\partial X_m\partial X_n},
{\partial^2\lad{\bf h}^k_0\rad\over\partial X_m\partial X_n},$$
$${\partial\over\partial X_m}(\lad{\bf v}^k_0\rad\lad{\bf v}^n_0\rad),
{\partial\over\partial X_m}(\lad{\bf v}^k_0\rad\lad{\bf h}^n_0\rad),
{\partial\over\partial X_m}(\lad{\bf h}^k_0\rad\lad{\bf h}^n_0\rad),$$
$$\lad{\bf v}^k_0\rad\lad{\bf v}^m_0\rad\lad{\bf v}^n_0\rad,
\lad{\bf h}^k_0\rad\lad{\bf v}^m_0\rad\lad{\bf v}^n_0\rad,
\lad{\bf h}^k_0\rad\lad{\bf h}^m_0\rad\lad{\bf v}^n_0\rad,
\lad{\bf h}^k_0\rad\lad{\bf h}^m_0\rad\lad{\bf h}^n_0\rad.$$
Коэффициенты перед этими членами являются решениями соответствующих
вспомогательных задач. Поскольку они не дают вклада в уравнения для
средних полей, эти вспомогательные задачи здесь не рассматриваются.

\mi{\bf 8. Уравнения порядка $\varepsilon^3$.}
Уравнения, полученные из членов рядов (5.1)-(5.3) порядка $\varepsilon^3$,
имеют вид
$${\cal L}^\omega(\bom_3,{\bf v}_3,{\bf h}_3,\theta_3)-{\partial\bom_1\over\partial T}
+\nu(2(\nabla_{\bf x}\cdot\nabla_{\bf X})\lbd\bom_2\rbd_v+\Delta_{\bf X}\bom_1)$$
$$+\nabla_{\bf X}\times({\bf V}\times\bom_2+{\bf v}_2\times\bOm
-{\bf H}\times(\nabla_{\bf X}\times{\bf h}_1+
\nabla_{\bf x}\times{\bf h}_2)$$
$$-{\bf h}_2\times(\nabla_{\bf x}\times{\bf H})
+{\bf v}_1\times\bom_0+{\bf v}_0\times\bom_1$$
$$-{\bf h}_1\times(\nabla_{\bf x}\times{\bf h}_0)
-{\bf h}_0\times(\nabla_{\bf x}\times{\bf h}_1+\nabla_{\bf X}\times{\bf h}_0))$$
$$+\nabla_{\bf x}\times(-{\bf H}\times(\nabla_{\bf X}\times{\bf h}_2)
+{\bf v}_2\times\bom_0+{\bf v}_1\times\bom_1+{\bf v}_0\times\bom_2$$
$$-{\bf h}_2\times(\nabla_{\bf x}\times{\bf h}_0)
-{\bf h}_1\times(\nabla_{\bf x}\times{\bf h}_1+\nabla_{\bf X}\times{\bf h}_0)$$
$$-{\bf h}_0\times(\nabla_{\bf x}\times{\bf h}_2+\nabla_{\bf X}\times{\bf h}_1))
+\beta\nabla_{\bf X}\theta_2\times{\bf e}_3=0;\eqn{39.1}$$
$${\cal L}^h({\bf v}_3,{\bf h}_3)-{\partial{\bf h}_1\over\partial T}
+\eta(2(\nabla_{\bf x}\cdot\nabla_{\bf X})\lbd{\bf h}_2\rbd_h+\Delta_{\bf X}{\bf h}_1)$$
$$+\nabla_{\bf X}\times({\bf v}_2\times{\bf H}+{\bf V}\times{\bf h}_2
+{\bf v}_1\times{\bf h}_0+{\bf v}_0\times{\bf h}_1)$$
$$+\nabla_{\bf x}\times({\bf v}_2\times{\bf h}_0
+{\bf v}_1\times{\bf h}_1+{\bf v}_0\times{\bf h}_2)=0;\eqn{39.2}$$
$${\cal L}^\theta({\bf v}_3,\theta_3)-{\partial\theta_1\over\partial T}
+\kappa(2(\nabla_{\bf x}\cdot\nabla_{\bf X})\theta_2+\Delta_{\bf X}\theta_1)
-({\bf v}_1\cdot\nabla_{\bf X})\theta_0-({\bf v}_0\cdot\nabla_{\bf X})\theta_1$$
$$-({\bf V}\cdot\nabla_{\bf X})\theta_2-({\bf v}_2\cdot\nabla_{\bf x})\theta_0
-({\bf v}_1\cdot\nabla_{\bf x})\theta_1-({\bf v}_0\cdot\nabla_{\bf x})\theta_2=0.\eqn{39.3}$$
Усреднение вертикальной компоненты (39.1) по быстрым прост\-ранственным
пере\-менным с учетом (7.1), (12.1)-(12.3), (13) при $n=0,1,2$,
краевых условий для $\bf V$, $\bf H$, ${\bf v}_i$ и ${\bf h}_i$
и соленоидальности ${\bf v}_0$ и ${\bf h}_0$ дает
$$-{\partial\la\bom_3\ra_v\over\partial t}-{\partial\la\bom_1\ra_v\over\partial T}
+\nu\Delta_{\bf X}\la\bom_1\ra_v$$
$$+\nabla_{\bf X}\times\la{\bf V}\times(\nabla_{\bf X}\times{\bf v}_1)
-{\bf V}\nabla_{\bf X}\cdot\lbd{\bf v}_1\rbd_h
-{\bf H}\times(\nabla_{\bf X}\times{\bf h}_1)
+{\bf H}\nabla_{\bf X}\cdot\lbd{\bf h}_1\rbd_h$$
$$+{\bf v}_0\times(\nabla_{\bf X}\times{\bf v}_0)
-{\bf v}_0\nabla_{\bf X}\cdot\lbd{\bf v}_0\rbd_h
-{\bf h}_0\times(\nabla_{\bf X}\times{\bf h}_0)
+{\bf h}_0\nabla_{\bf X}\cdot\lbd{\bf h}_0\rbd_h\ra_h=0.$$
Усредняя это уравнение по быстрому времени с учетом экспоненциального убы\-вания
$\bxi_0$ и $\bxi_1$, подставляя представления (16.2), (16.3), (23.2) и (23.3)
и используя (14) при $n=1$ и равенство нулю пространственных средних
горизонтальных компонент подынтегральных выражений в (35), получим
$$\nabla_{\bf X}\times\left[-{\partial\over\partial T}\lad{\bf v}_0\rad_h
+\nu\Delta_{\bf X}\lad{\bf v}_0\rad_h\right.
+\sum_{j=1}^2\sum_{m=1}^2\sum_{k=1}^2
\left({\bf D}^{v,v}_{m,k,j}{\partial^2\lad{\bf v}_0^k\rad\over\partial X_j\partial X_m}
+{\bf D}^{h,v}_{m,k,j}{\partial^2\lad{\bf h}_0^k\rad\over\partial X_j\partial X_m}\right)$$
$$+(\lad{\bf v}_0\rad_h\cdot\nabla_{\bf X})\lad{\bf v}_0\rad_h
-(\lad{\bf h}_0\rad_h\cdot\nabla_{\bf X})\lad{\bf h}_0\rad_h$$
$$+\sum_{j=1}^2\sum_{m=1}^2\sum_{k=1}^2{\partial\over\partial X_j}\left(
{\bf A}^{vv,v}_{m,k,j}\lad{\bf v}^k_0\rad\lad{\bf v}^m_0\rad
+{\bf A}^{vh,v}_{m,k,j}\lad{\bf v}^k_0\rad\lad{\bf h}^m_0\rad
\left.+{\bf A}^{hh,v}_{m,k,j}\lad{\bf h}^k_0\rad\lad{\bf h}^m_0\rad\right)\right]=0.\eqn{40}$$
Здесь $\bf D$ обозначают коэффициенты членов уравнения для среднего потока,
отве\-чающие т.н. вихревой коррекции кинематической вязкости, $\bf A$ -- квадратичных
членов, отвечающие т.н. вихревой коррекции кинематической адвекции:
$${\bf D}^{v,v}_{m,k,j}=\lad
-{\bf V}^j{\bf G}^{v,v}_{m,k}-{\bf V}({\bf G}^{v,v}_{m,k})^j
+{\bf H}^j{\bf G}^{v,h}_{m,k}+{\bf H}({\bf G}^{v,h}_{m,k})^j\rad_h,\eqn{41.1}$$
$${\bf D}^{h,v}_{m,k,j}=\lad
-{\bf V}^j{\bf G}^{h,v}_{m,k}-{\bf V}({\bf G}^{h,v}_{m,k})^j
+{\bf H}^j{\bf G}^{h,h}_{m,k}+{\bf H}({\bf G}^{h,h}_{m,k})^j\rad_h,\eqn{41.2}$$
$${\bf A}^{vv,v}_{m,k,j}=\lad
-{\bf V}^j{\bf Q}^{vv,v}_{m,k}-{\bf V}({\bf Q}^{vv,v}_{m,k})^j
+{\bf H}^j{\bf Q}^{vv,h}_{m,k}+{\bf H}({\bf Q}^{vv,h}_{m,k})^j$$
$$+({\bf S}^{v,v}_k)^j{\bf S}^{v,v}_m-({\bf S}^{v,h}_k)^j{\bf S}^{v,h}_m\rad_h,\eqn{42.1}$$
$${\bf A}^{vh,v}_{m,k,j}=\lad
-{\bf V}^j{\bf Q}^{vh,v}_{m,k}-{\bf V}({\bf Q}^{vh,v}_{m,k})^j
+{\bf H}^j{\bf Q}^{vh,h}_{m,k}+{\bf H}({\bf Q}^{vh,h}_{m,k})^j$$
$$+({\bf S}^{v,v}_k)^j{\bf S}^{h,v}_m+({\bf S}^{h,v}_m)^j{\bf S}^{v,v}_k
-({\bf S}^{v,h}_k)^j{\bf S}^{v,h}_m-({\bf S}^{h,h}_m)^j{\bf S}^{v,h}_k\rad_h,\eqn{42.2}$$
$${\bf A}^{hh,v}_{m,k,j}=\lad
-{\bf V}^j{\bf Q}^{hh,v}_{m,k}-{\bf V}({\bf Q}^{hh,v}_{m,k})^j
+{\bf H}^j{\bf Q}^{hh,h}_{m,k}+{\bf H}({\bf Q}^{hh,h}_{m,k})^j$$
$$+({\bf S}^{h,v}_k)^j{\bf S}^{h,v}_m-({\bf S}^{h,h}_k)^j{\bf S}^{h,h}_m\rad_h.\eqn{42.3}$$

(36) и (40) представляют из себя замкнутую систему уравнений для средних полей
(точнее, главных членов разложений средних полей в ряд по отношению пространственных
масштабов). В (36) только горизонтальные компоненты не равны тождественно нулю,
а в (40) -- только вертикальная компонента. От (40) можно перейти к уравнению
без внешнего ротора, если ввести длинномасштабное давление; по форме оно не
отличается от уравнения, выведенного Желиговским [2006]. Однако, поскольку
$\lad{\bf v}_0\rad_h$ и $\lad{\bf h}_0\rad_h$ соленоидальны относительно
медленных переменных, естественнее решать систему уравнений в терминах функции
тока для потока $\lad{\bf v}_0\rad_h$ и аналогичного векторного потенциала для
$\lad{\bf h}_0\rad_h$, рас\-сматривая вертикальную компоненту (40) без изменений,
и взяв ротор (36) и рассматривая вертикальную -- единственную ненулевую --
компоненту полученного уравнения.

Отметим, что сила Кориолиса входит только в формулировку вспомогатель\-ных задач
и отсутствует в уравнениях для средних полей. Однако, вывод этих уравнений
не изменяется, если дополнительно имеет место вращение жидкости
в медленном времени, и (полный) вектор угловой скорости имеет вид
$\tau{\bf e}_3+\varepsilon^2\btau$. Тогда в усредненном уравнении
(40) появляется слагаемое, соответствующее силе Кориолиса при вращении слоя
жидкости с угловой скоростью $\btau$.

\mi{\bf 8. Вычисление коэффициентов вихревых членов.}
Для вычисления коэф\-фициентов $\bf A$ и $\bf D$ в уравнениях (36) и (40) достаточно
решить 22 вспомогательные задачи (4 первого, 8 второго и 10 третьего типа).
Однако можно воспользоваться
тем обстоятельством, что решения вспомогательных задач второго
$({\bf G}^{\cdot,v}_{m,k}$, ${\bf G}^{\cdot,v}_{m,h})$ и третьего
$({\bf Q}^{\cdot\cdot,v}_{m,k},{\bf Q}^{\cdot\cdot,h}_{m,k})$
типов входят в средние (37) и (38) только в качестве сомножителей скалярного
произведения с вектором $({\bf H\times e}_3,-{\bf V\times e}_3,0)$, а в (41) и (42) --
с векторами ${\bf W}_{n,j}\equiv(-{\bf V}^j{\bf e}_n-{\bf V}^n{\bf e}_j,{\bf H}^j{\bf e}_n+{\bf H}^n{\bf e}_j,0)$
(4 вектора с\break$n,j=1,2$). (Здесь используется скалярное произведение
$\lad{\bgamma\cdot\bf W}\rad$, где $\bgamma$ и $\bf W$ -- 7-мерные векторные поля,
заданные в слое.) Более того, поскольку в (40) градиенты под знаком ротора несущественны,
в (41) и (42) достаточно вычислять скалярные произведения с ${\bf W}_{1,2}$,
${\bf W}_{2,1}$ и ${\bf W}_{1,1}-{\bf W}_{2,2}$.
Поэтому число вспомогатель\-ных задач, которые необходимо
решить, можно уменьшить до 8, если ввести в рассмотрение (как в работах
[Zheligovsky, 2005], [Желиговский, 2006]) {\it вспомога\-тельные задачи
для сопряженного оператора} (такой же вычислительной слож\-ности, как и
рассмотренные выше вспомогательные задачи).

Именно, пусть
$${\bf W}^v=\hat{\bf W}^v+\nabla W^v,\quad {\bf W}^h=\hat{\bf W}^h+\nabla W^h,\quad
\bgamma^v=\hat{\bgamma}^v+\nabla\gamma^v,\quad\bgamma^h=\hat{\bgamma}^h+\nabla\gamma^h$$
(все дифференциальные операторы -- в быстрых переменных)  --
разложения 3-мерных векторных компонент $\bf W$ и $\bgamma$ на соленоидальную и
потенциальную состав\-ляющие, $\hat{\bgamma}$ и $\bf Z$ удовлетворяют системам уравнений
$${\cal L}^v(\hat{\bgamma}^v,\hat{\bgamma}^h,\hat{\gamma}^\theta,\hat{\gamma}^p)=\hat{\bf F}^v,\quad
{\cal L}^h(\hat{\bgamma}^v,\hat{\bgamma}^h)=\hat{\bf F}^h,\quad
{\cal L}^\theta(\hat{\bgamma}^v,\hat{\gamma}^\theta)=\hat{F}^\theta,\eqn{43.1}$$
$$\nabla\cdot\hat{\bgamma}^v=\nabla\cdot\hat{\bgamma}^h=0;\eqn{43.2}$$
$$({\cal L}^*)^v({\bf Z}^v,{\bf Z}^h,Z^\theta)=\hat{\bf W}^v,\quad
({\cal L}^*)^h({\bf Z}^v,{\bf Z}^h)=\hat{\bf W}^h,\quad
({\cal L}^*)^\theta({\bf Z}^v,Z^\theta)=0,\eqn{44.1}$$
$$\nabla\cdot{\bf Z}^v=\nabla\cdot{\bf Z}^h=0\eqn{44.2}$$
и соответствующим краевым условиям. Тогда
\pagebreak
$$\lad\bgamma^v\cdot{\bf W}^v+\bgamma^h\cdot{\bf W}^h\rad
=\lad\nabla W^v\cdot\nabla\gamma^v+\nabla W^h\cdot\nabla\gamma^h+
\hat{\bf W}^v\cdot\hat{\bgamma}^v+\hat{\bf W}^h\cdot\hat{\bgamma}^h\rad$$
$$=\lad\nabla W^v\cdot\nabla\gamma^v+\nabla W^h\cdot\nabla\gamma^h$$
$$+(\hat{\bgamma}^v,\hat{\bgamma}^h,\hat{\gamma}^\theta)
\cdot\left(({\cal L}^*)^v({\bf Z}^v,{\bf Z}^h,Z^\theta),
({\cal L}^*)^h({\bf Z}^v,{\bf Z}^h),({\cal L}^*)^\theta({\bf Z}^v,Z^\theta)\right)\rad$$
$$=\lad\nabla W^v\cdot\nabla\gamma^v+\nabla W^h\cdot\nabla\gamma^h+
\hat{\bf F}^v\cdot{\bf Z}^v+\hat{\bf F}^h\cdot{\bf Z}^h+\hat{F}^\theta Z^\theta\rad.\eqn{45}$$

Тем самым, решение вспомогательных задач второго и третье\-го типов оказы\-вается
возможным избежать -- достаточно знать соответствующие правые части в
определяющих их системах уравнений. Для этого
при вычислении коэффициентов (37) и (41) необходимо
воспользоваться условиями (24.4), (24.6), (25.4), (25.6) для вычисления
потенциальных составляющих полей ${\bf G}^{\cdot,v}_{m,k}$ и
${\bf G}^{\cdot,h}_{m,k}$ и для перехода к их соленоидальным составляющим
в (24.5), (25.5), (24.7) и (25.7),  и в дополнение использовать (24.2) и (25.2)
для перехода к соленоидальным составляющим ${\bf G}^{\cdot,\cdot}_{m,k}$
в (24.1) и (25.1). Аналоги (24.1) и (25.1) для соленоидальных неизвестных
можно привести к виду (43.1), рассматривая их векторный потенциал.
Уравнения (26.1), (27.1) и (28.1) используются в форме ($26.1'$), ($27.1'$)
и ($28.1'$), соответственно.

В (44) ${\cal L}^*=(({\cal L}^*)^v,\,({\cal L}^*)^h,\,({\cal L}^*)^\theta)$
-- оператор, формально сопряженный к\break${\cal L}=({\cal L}^v,\,{\cal L}^h,\,{\cal L}^\theta)$
относительно рассматриваемого скалярного произведения:
$$({\cal L}^*)^v({\bf v,h},\theta)\equiv{\partial{\bf v}\over\partial t}+\nu\Delta{\bf v}
-\nabla\times({\bf V}\times{\bf v})+{\cal P}({\bf H}\times(\nabla\times{\bf h})
-{\bf v}\times(\nabla\times{\bf V})$$
$$-\tau{\bf v\times e}_3+\delta\theta{\bf e}_3+\theta\nabla\Theta),$$
$$({\cal L}^*)^h({\bf v,h})\equiv{\partial{\bf h}\over\partial t}+\eta\Delta{\bf h}
+\nabla\times({\bf H}\times{\bf v})+{\cal P}({\bf v}\times(\nabla\times{\bf H})
-{\bf V}\times(\nabla\times{\bf h})),$$
$$({\cal L}^*)^\theta({\bf v},\theta)\equiv{\partial\theta\over\partial t}
+\kappa\Delta\theta+({\bf V}\cdot\nabla)\theta+\beta{\bf v}^3,$$
$\cal P$ -- оператор
проекции трехмерного векторного поля в пространст\-во соленои\-дальных полей.
Оператор ${\cal L}^*$ определен корректно, например, если исходное состояние
${\bf V,H},\theta$ периодично по пространству и стационарно или периодично
по времени, и область его определения состоит из функций, также стационарных
и/или периодичных. В общем случае численное решение задачи (44)
проблема\-тично, т.к. оператор ${\cal L}^*$ не параболический. Обойти эту сложность можно
следую\-щим образом: для неизвестных полей ${\bf Z}^{\cdot,\cdot}$ нулевые
"начальные"\ условия ставим при $t=\hat{t}>0$, а затем решаем (44)
в сторону убывающего времени до $t=0$ (при обращении времени
операторы $({\cal L}^*)^v,\,({\cal L}^*)^h$ и $({\cal L}^*)^\theta$
становятся параболическими). Полученное решение,
зависящее от $\hat{t}$, обозначим ${\bf Z}(\hat{t};{\bf x},t)$. Тогда
пространственно-временн\'ое усреднение по быстрым переменным скалярных
произведений в (45) определяется равенством
$$\lad{\bf F\cdot Z}\rad=\lim_{\hat{t}\to\infty}
\lim_{\ell\to\infty}{1\over\hat{t}L\ell^2}\int_0^{\hat{t}}\int_{-L/2}^{L/2}
\int_{-\ell/2}^{\ell/2}\int_{-\ell/2}^{\ell/2}
{\bf F}({\bf x},t)\cdot{\bf Z}(\hat{t};{\bf x},t)\,dx_1\,dx_2\,dx_3\,dt$$
(если указанный предел существует, -- например, если
исходное состояние ${\bf V,H},P$ квазипериодично по времени).

\pagebreak
\mi{\bf 8. Выводы.}
Рассмотрена устойчивость центрально-симметрич\-ного МГД состо\-яния,
не имеющего больших масштабов, по отношению к возмущению, в котором
присутствуют большие простран\-ственные и временн\'ые масштабы.
Построено асимптотическое разложение возмущения и выведены уравнения (36), (40)
эволю\-ции в нелинейном режиме главного члена разложения возмущения, усредненного по малым
пространственно-временн\'ым масштабам.

\mi{\bf Благодарности}.
Работа частично финансировалась РФФИ (грант 04-05-64699).

\bigskip
\mi{\bf Литература}

\mi{\it С.И.Брагинский.} О самовозбуждении магнитного поля при движении
хорошо проводящей жидкости // ЖЭТФ. 1964а. Т.~47. С.~1084-1098.

\mi{\it С.И.Брагинский.} К теории гидромагнитного динамо // ЖЭТФ. 1964б.
Т.~47. С.~2178-2193.

\mi{\it С.И.Брагинский.} Кинематические модели гидромагнитного динамо Земли //
Гео\-магн. аэроном. 1964в. Т.~4, $N$ 4. С.~732-747.

\mi{\it С.И.Брагинский.} Магнитогидродинамика Земного ядра //
Геомагн. аэроном. 1964д. Т.~4, $N$ 5. С.~898-916.

\mi{\it С.И.Брагинский.} Магнитные волны в ядре Земли //
Геомагн. аэроном. 1967. Т.~7, $N$ 6. С.~1050-1060.

\mi{\it С.И.Брагинский.} Почти аксиально-симметричная модель гидромагнитного
дина\-мо Земли. I. // Геомагн. аэроном. 1975. Т.~15, $N$ 1. С.~149-156.

\mi{\it В.А.Желиговский.} О линейной устойчивости стационарных
прост\-ранственно-периодических магнитогидродинамических систем
к длиннопериодным возмуще\-ниям // Физика Земли. 2003. $N$ 5. C.~65-74
[http://arxiv.org/abs/nlin/0512076].

\mi{\it В.А.Желиговский.} Слабо нелинейная устойчивость магнитогидродинамических
систем, имеющих центр симметрии, к возмущениям с большими масштабами //
Физика Земли. 2006. $N$ 3 [http://arxiv.org/abs/nlin/0601012].

\mi{\it Г.Моффат.} Возбуждение магнитного поля в проводящей среде.
М.: Мир. 1980. 340 с.

\mi{\it Е.Паркер.} Космические магнитные поля. Т. 1, 608 с.
Т. 2, 480 с. М.: Мир. 1982.

\mi{\it У.Паркинсон.} Введение в геомагнетизм. М.: Мир. 1986. 528 с.

\mi{\it Baptista M., Gama S.M.A., Zheligovsky V.}
Multiple-scale \hbox{expansions} for incompressible MHD systems.
Preprint 2004-11, Centro de Matem\'a\-tica da Universidade do Porto,
Faculdade de Ci\^encias da Universidade do Porto, 2004\break
[http://cmup.fc.up.pt/cmup/preprints/2004-11.pdf].

\mi{\it Baptista M., Gama S.M.A., Zheligovsky V.}
Eddy diffusivity in convective hydromagnetic systems.
Подано в Physical Review E, 2005\break
[http://xxx.lanl.gov/abs/nlin.CD/0511020].

\pagebreak
\mi{\it Bensoussan A., Lions J.-L., Papanicolaou G.} Asymptotic
analysis for periodic structures. Amsterdam: North Holland. 1978. 700 pp.

\mi{\it Busse F.H.} Homogenous dynamos in planetary cores and in the laboratory
// Ann. Rev. Fluid Mech. 2000. V.~32. P.~383-408.

\mi{\it Cioranescu D., Donato P.} An introduction to homogenization. Oxford
Univ. Press. 1999. 262 pp.

\mi{\it Christensen U., Olson P., Glatzmaier G.A.} Numerical modeling
of the geodynamo: A systematic parameter study // Geophys. J. Int. 1999.
V.~138. P.~393-409.

\mi{\it Christensen U.R.} Mantle rheology, constitution, and convection.
/ Peltier W.R. (Ed.), Mantle convection. Plate tectonics and global
dynamics. New York: Gordon and Breach. 1989. P.~595-656.

\mi{\it Cross M.C., Newell A.C.} Convection patterns in large aspect ratio
systems // Physica D. 1984. V.~10. P.~299-328.

\mi{\it Frisch U., Legras B., Villone B.} Large-scale Kolmogorov flow on the
beta-plane and resonant wave interactions // Physica D. 1996. V.~94. P.~36-56.

\mi{\it Gama S., Chaves M.} Time evolution of the eddy viscosity in
two-dimensional Navier-Stokes flow. Phys. Rev. Lett. 2000. V.~61. P.~2118-2120.

\mi{\it Gama S., Vergassola M., Frisch U.} Negative eddy viscosity in
isotropically forced two-dimensional flow: linear and nonlinear dynamics //
J. Fluid Mech. 1994. V.~260. P.~95-126.

\mi{\it Glatzmaier G.A., Roberts P.H.} A three-dimensional convective dynamo
solution with rotating and finitely conducting inner core and mantle //
Phys.~Earth Planet.~Inter. 1995а. V.~91. P.~63-75.

\mi{\it Glatzmaier G.A., Roberts P.H.} A three-dimensional self-consistent
computer simulation of a geomagnetic field reversal //
Nature. 1995б. V.~377. P.~203-209.

\mi{\it Glatzmaier G.A., Roberts P.H.} An anelastic geodynamo simulation
driven by compositional and thermal convection // Physica D. 1996а. V.~97.
P.~81-94.

\mi{\it Glatzmaier G.A., Roberts P.H.} Rotation and magnetism of Earth's
inner core // Science. 1996б. V.~274. P.~1887-1891.

\mi{\it Glatzmaier G.A., Roberts P.H.} Simulating the geodynamo //
Contemporary physics. 1997а. V.~38. P.~269-288.

\mi{\it Glatzmaier G.A., Roberts P.H.} Computer simulations of the Earth's
magnetic field // Geowissenschaften. 1997б. V.~15. P.~95.

\mi{\it Glatzmaier G.A., Coe R.S., Hongre L., Roberts P.H.} The role
of the Earth's mantle in controlling the frequency of geomagnetic reversals //
Nature. 1999. V.~401. P.~885-890.

\pagebreak
\mi{\it Jikov V.V., Kozlov S.M., Oleinik O.A.} Homogenization of differential
operators and integral functionals. Berlin: Springer-Verlag. 1994. 570 pp.

\mi{\it Jones C.A.} Convection-driven geodynamo models //
Phil. Trans. R. Soc. Lond. 2000. V.~A358. P.~873-897.

\mi{\it Lanotte A., Noullez A., Vergassola M., Wirth A.}
Large-scale dynamo by negative magnetic eddy diffusivities.
Geophys.~Astrophys.~Fluid Dynam. 1999. V.~91. P.~131-146.

\mi{\it Merrill R.T., McEllhiny M.W., McFadden Ph.L.} The magnetic field of
the Earth. Paleomagnetism, the core and the deep mantle. San Diego:
Academic Press. 1996. 527 pp.

\mi{\it Newell A.C.} Two-dimensional convection patterns in large aspect
ratio systems / Fujita H. (Ed.), Nonlinear partial differential equations
in applied science. Amsterdam: North-Holland. 1983. P.~202-231.

\mi{\it Newell A.C., Passot T., Lega J.} Order parameter equations for patterns
// Ann. Rev. Fluid Mech. 1993. V.~25. P.~399-453.

\mi{\it Newell A.C., Passot T., Bowman C., Ercolani N., Indik R.}
Defects are weak and self-dual solutions of the Cross-Newell phase diffusion
equation for natyral patterns // Physica D. 1996. V.~97. P.~185-205.

\mi{\it Newell A.C., Passot T., Souli M.} Convection at finite Rayleigh numbers
in large-aspect-ratio containers // Phys. Rev. Lett. 1990a. V.~64 (20).
P.~2378-2381.

\mi{\it Newell A.C., Passot T., Souli M.} The phase diffusion and mean drift
equations for convection at finite Rayleigh numbers in large containers //
J. Fluid Mech. 1990б. V.~220. P.~187-552.

\mi{\it Oleinik O.A., Shamaev A.S., Yosifian G.A.} Mathematical problems
in elasticity and homogenization. Amsterdam: Elsevier Science Publishers. 1992.
398 pp.

\mi{\it Olson P., Christensen U., Glatzmaier G.A.} Numerical modeling of
the geodynamo: mechanisms of field generation and equilibration //
J. Geophys. Res. 1999. V.~104. P.~10383-10404.

\mi{\it Peltier W.R.} Mantle viscosity / Peltier W.R. (Ed.), Mantle convection.
Plate tectonics and global dynamics. New York: Gordon and Breach. 1989. P.~389-478.

\mi{\it Ponty Y., Passot T., Sulem P.L.} Pattern dynamics in rotating convection
at finite Prandtl number // Phys. Rev. E. 1997. V.~56 (4). P.~4162-4178.

\mi{\it Ponty Y., Gilbert A.D., Soward A.M.} Kinematic dynamo action
in flows driven by shear and convection. // J. Fluid Mech. 2001а. V.~435.
P.~261-287.

\pagebreak
\mi{\it Ponty Y., Gilbert A.D., Soward A.M.} Dynamo action due to Ekman layer
instability / Chossat P., Armbruster D., Oprea I. (Eds.), Dynamo and dynamics,
a mathematical challenge. Boston: Kluwer. 2001б. P.~75-82.

\mi{\it Ponty Y., Gilbert A.D., Soward A.M.} The onset of thermal convection in
Ekman--Couette shear flow with oblique rotation // J. Fluid Mech. 2003. V.~487.
P.~91-123.

\mi{\it Roberts P.H., Glatzmaier G.A.} The geodynamo, past, present and
future // Geophys. Astrophys. Fluid Dynam. 2001. V.~94. P.~47-84.

\mi{\it Rotvig J., Jones C.A.} Rotating convection driven dynamos at low Ekman
number // Phys. Rev. E. 2002. V.~66 056308. P.~1-15\break
[http://link.aps.org/abstract/PRE/v66/e056308].

\mi{\it Sagaut P.} Large eddy simulation for incomressible flows. Berlin:
Springer-Verlag. 2006. 556 pp.

\mi{\it Sarson G.R., Jones C.A.} A convection driven geodynamo reversal model //
Phys. Earth Planet. Inter. 1999. V.~111. P.~3-20.

\mi{\it Soward A.M.} A kinematic theory of large magnetic Reynolds number
dynamos // Phil. Trans. Roy. Soc. A. 1972. V.~272. P.~431-462.

\mi{\it Soward A.M.} A convection driven dynamo I. The weak field case //
Phil. Trans. Roy. Soc. A. 1974. V.~275. P.~611-651.

\mi{\it Starr V.P.} Physics of negative viscosity phenomena. New York:
McGraw-Hill. 1968.

\mi {\it Wirth A., Gama S., Frisch U.} Eddy viscosity of
three-dimensional flow // J. Fluid Mech. 1995. V.~288. P.~249-264.

\mi{\it Zeldovich Ya.B., Ruzmaikin A.A., Sokoloff D.D.} Magnetic
fields in astrophysics. New York: Gordon and Breach. 1983. 365 pp.

\mi{\it Zhang K., Jones C.A.} The effect of hyperviscosity on geodynamo models
// Geophys. Res. Lett. 1997. V.~24. P.~2869-2872.

\mi{\it Zhang K., Schubert G.} Magnetohydrodynamics in rapidly rotating
spherical systems // Ann. Rev. Fluid Mech. 2000. V.~32. P.~409-443.

\mi{\it Zheligovsky V.A., Podvigina O.M., Frisch U.} Dynamo effect
in parity-invariant flow with large and moderate separation of scales //
Geophys. Astrophys. Fluid Dynam. 2001. V.~95. P.~227-268
[http://xxx.lanl.gov/abs/nlin.CD/0012005].

\mi{\it Zheligovsky V.A., Podvigina O.M.} Generation of multiscale magnetic
field by parity-invariant time-periodic flows // Geophys. Astrophys. Fluid
Dynam. 2003. V.~97. P.~225-248 [http://xxx.lanl.gov/abs/physics/0207112].

\mi{\it Zheligovsky V.A.} Convective plan-form two-scale dynamos in a plane
layer // Geophys. Astrophys. Fluid Dynam. 2005. V.~99. P.~151-175\break
[http://arxiv.org/abs/physics/0405045].
\end{document}